%% file: aa.tex
\documentclass{aa}  

\usepackage{graphicx}
\usepackage{txfonts}
\usepackage{multirow}
\usepackage{hyperref}
\usepackage{breakurl}
\newcommand{\chandra}{{\it Chandra}}
\newcommand{\swift}{{\it Swift}}
\newcommand{\xmm}{{\it XMM-Newton}}
\newcommand{\asat}{{\it AstroSat}}
\newcommand{\nicer}{{\it NICER}}

\begin{document}

   \title{The Super-Soft Source Phase of the recurrent nova V3890\,Sgr}

%   \subtitle{The Super-Soft Source Phase of V3890 Sgr}

   \author{J.-U. Ness\inst{\ref{esa}}\and
  A.P. Beardmore\inst{\ref{leicester}}\and
  P. Bezak\inst{\ref{slovak}}\and
  A. Dobrotka\inst{\ref{slovak}}\and
  J.J. Drake\inst{\ref{sao}}\and
  B. Vander Meulen\inst{\ref{ugent}}\and
  J.P. Osborne\inst{\ref{leicester}}\and
  M. Orio\inst{\ref{wisconsin},\ref{padova}}\and
  K.L. Page\inst{\ref{leicester}}\and
  C. Pinto\inst{\ref{palermo}}\and
  K.P. Singh\inst{\ref{singh}}\and
  S. Starrfield\inst{\ref{asu}}
    }

\institute{European Space Agency (ESA), European Space Astronomy Centre (ESAC), Camino Bajo del Castillo s/n, 28692 Villanueva de la Ca\~nada, Madrid, Spain; corresponding author: \email{jan.uwe.ness@esa.int}\label{esa}
\and
School of Physics \& Astronomy, University of Leicester, Leicester, LE1 7RH, UK\label{leicester}
        \and
Advanced Technologies Research Institute, Faculty of Materials Science and Technology in Trnava, Slovak University of Technology in Bratislava, Bottova 25, 917 24 Trnava, Slovakia\label{slovak}
     \and
Harvard-Smithsonian Center for Astrophysics, 60 Garden Street, Cambridge, MA 02138, USA\label{sao}
        \and
Sterrenkundig Observatorium, Ghent University, Krijgslaan 281 - S9, 9000 Gent, Belgium\label{ugent}
      \and
   Dept. of Astronomy, University of Wisconsin, 475 N, Charter Str., Madison, WI 53706, USA\label{wisconsin}
     \and
     INAF-Osservatorio di Padova, Vicolo Osservatorio 5, 35122 Padova, Italy\label{padova}
     \and
%Observatories of the Carnegie Institution of Science, 813 Santa Barbara Street, Pasadena, CA 91101, USA\label{carnegie}
%     \and
       INAF-IASF Palermo, Via U. La Malfa 153, I-90146 Palermo, Italy\label{palermo}
      \and
Indian institute of Science Education and Research Mohali, Sector 81, SAS Nagar, Manauli PO, 140306, India\label{singh}
      \and
    School of Earth and Space Exploration, Arizona State University, Tempe, AZ 85287-1404, USA\label{asu}
}
   \authorrunning{Ness et al.}
   \titlerunning{SSS grating spectrum of V3890\,Sgr}
   \date{Received \today; accepted }

% \abstract{}{}{}{}{} 
% 5 {} token are mandatory
 
  \abstract
% Context
{
The 30-year recurrent symbiotic nova \object{V3890 Sgr} exploded 2019 August 28
and was observed with multiple X-ray telescopes. \swift\ and \asat\
monitoring revealed slowly declining hard X-ray emission from
shocks between the nova ejecta and the stellar wind of the companion.
Later, highly variable Super-Soft-Source (SSS) emission was seen.
An \xmm\thanks{\xmm\ is an ESA science mission with
   instruments and
   contributions directly funded by ESA Member States and NASA.}
observation during the SSS phase captured the high degree of X-ray
variability in terms of a deep dip in the middle of the observation.
}
{
This observation adds to the growing sample of diverse SSS
spectra and allows spectral comparison of low- and high-state emission
to identify the origin of variations and subsequent effects of such
dips, all leading to new insights into how the nova ejecta evolve.
}
{
Based on initial visual inspection, quantitative modelling approaches
were conceptualised to test hypotheses of interpretation. The light
curve is analysed with a power spectrum analysis before and after
the dip and an eclipse model to test the hypothesis of occulting
clumps like in \object{U Sco}. A phenomenological spectral model (SPEX) is used to
fit the complex RGS spectrum accounting for all known atomic physics.
A blackbody source function is assumed like in all atmosphere radiation
transport models while the complex radiation transport processes are
not modelled. Instead, one or multiple absorbing layers are used to model
the absorption lines and edges, taking into account all state of the art
knowledge of atomic physics.
   }
   {
In addition to the central deep dip, there is an initial rise of similar
depth and shape and, after the deep dip, there are smaller dips of
$\sim 10$\% amplitude, which might be periodic over 18.1-minutes.
Our eclipse model of the dips yields clump sizes and orbital radii of
0.5-8 and 5-150 white dwarf radii, respectively. The simultaneous
XMM-Newton UV light curve shows no significant variations beyond slow fading.
The RGS spectrum contains both residual shock emission at short
wavelengths and the SSS emission at longer wavelengths. The shock
temperature has clearly decreased compared to an earlier \chandra\
observation (day 6). The dip spectrum is dominated by emission lines
like in U\,Sco. The intensity of underlying blackbody-like
emission is much lower with the blackbody normalisation yielding
a similar radius as during the brighter phases, while the lower
bolometric luminosity is ascribed to lower $T_{\rm eff}$. This would
be inconsistent with clump occultations unless Compton scattering of
the continuum emission reduces the photon energies to mimic a lower
effective temperature. However, systematic uncertainties are high.
The absorption lines in the bright SSS spectrum are blue-shifted by
$870\pm10$\,km\,s$^{-1}$ before the dip and slightly faster,
$900\pm10$\,km\,s$^{-1}$, after the dip. The reproduction of the
observed spectrum is astonishing, especially that only a single absorbing
layer is necessary while three such layers are needed to reproduce
the RGS spectrum of V2491\,Cyg. The ejecta of V3890\,Sgr are thus more
homogeneous than many other SSS spectra indicate.
Abundance determination is in principle possible but highly uncertain.
Generally, solar abundances are found except for N and possibly O higher
by an order of magnitude.
   }
   {
High-amplitude variability of SSS emission can be explained in several
ways without having to give up the concept of constant bolometric luminosity.
Variations in photospheric radius can expose deeper-lying plasma that could
pulse with 18.1 minutes and that would yield a higher outflow velocity. Also,
clump occultations are consistent with the observations.
    }

   \maketitle

%
%-------------------------------------------------------------------

\section{Introduction}

\begin{figure*}[!ht]
\resizebox{\hsize}{!}{\includegraphics{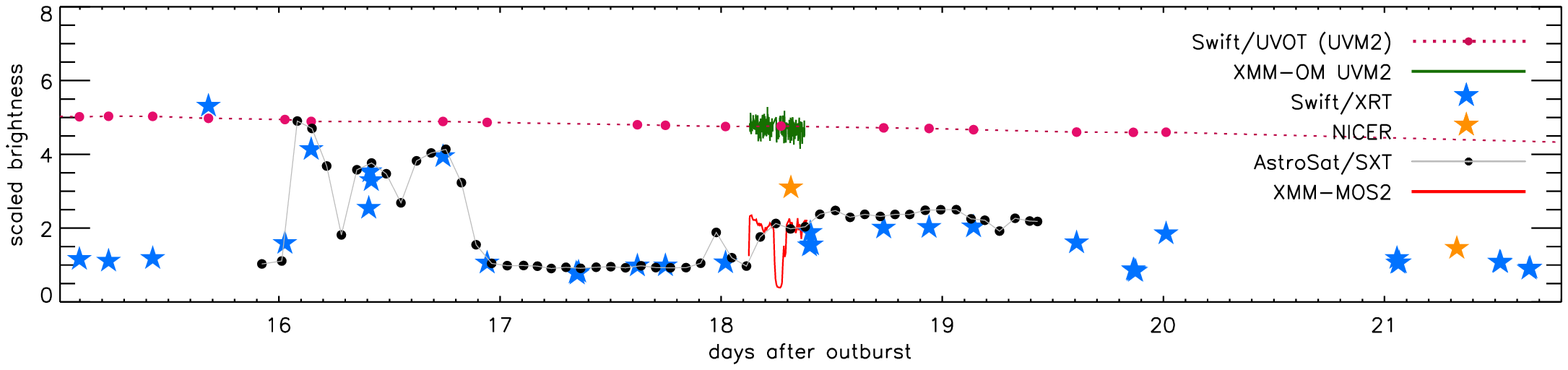}}%mmlc
\caption{\label{fig:mmlc}Multi-mission X-ray/UV light curves during the SSS
phase. UV data with UVM2 filter ($2120\pm500$\,\AA) were taken with
\swift/UVOT (ObsID ranges 00045788033-00045788053 and
00011558001-00011558021) and \xmm/OM (UV) and X-ray data over the energy range $\sim
0.3-10$\,keV with \asat/SXT (ObsIDs 9000003142 and 9000003160), \swift/XRT, \nicer\
(ObsIDs 2200810104 on day 18.3 and 2200810106 on day 21.3), and the MOS2 light
curve extracted from our new \xmm\ observation (see legend). The
\xmm\ MOS2 and OM/UVM2 light curves were taken between two \swift\
observations but overlap with 5 \asat/SXT observations. The high time
resolution of the \xmm\ observation demonstrates the high degree of
variability in X-rays on time scales shorter than the Earth occultation
gaps between \swift\ and \asat\ observations. Meanwhile, in UV, a slow
declining trend can be seen that is consistent in the \swift/UVOT and
\xmm/OM data. See \S\ref{sect:obs} for discussion.}
\end{figure*}

While there is a high degree of diversity among stars, binaries are
even more diverse, not only because of all types of combinations of
different stars but also various types of interactions between the
two binary components.
The class of Cataclysmic Variables (CVs) is defined by a white dwarf
(WD) primary that accretes material from a secondary star.
Properties of the WD that drive the observable characteristics are 
its mass, magnetic fields and, to a more subtle degree, parameters such
as its spin period and surface composition. Accreting WDs with low magnetic
fields are surrounded by an accretion disk while those with strong
magnetic fields, known as polars, accrete directly towards the magnetic
poles. The properties of the companion star determine the properties
of the accreted material whereas the properties of the orbit (e.g.
binary separation) drive parameters such as the accretion rate.\\

The class of {\em symbiotic binaries} is characterised by an evolved
secondary star revolving around a WD primary in a wide (years)
orbit. Evolved stars contribute a dense stellar wind with the
capacity to absorb significant amounts of kinetic energy via shocks.
While close CV binaries accrete via an overflow of material from the
companion star through the Lagrange point 1 (L1), in a wide
binary system, the companion star may not fill its Roche Lobe. However,
the high density of the stellar wind of an evolved star can still
facilitate a significant accretion rate (wind accretor).\\

The class of {\em novae} are those CVs that accumulate hydrogen-rich
accreted material on the surface of the white dwarf until the pressure
is high enough to ignite explosive thermo-nuclear runaway
reactions. The energy produced by CNO cycle nuclear fusion of hydrogen
to helium is initially released as high-energy radiation that couples
with the surface material that is then ejected via radiation pressure.
The ejected shell around the WD is optically thick to high-energy
radiation which is reprocessed to escape at optical and UV wavelengths
through a photosphere (with variable radius) after undergoing complex
radiation transport processes. Since the ejecta are not in hydrostatic
equilibrium, the photospheric radius shrinks with time as the density
of the outer ejecta decreases with the continuing expansion. This
allows successively deeper views into the outflow, and $T_{\rm eff}$
thus increases with time. Effectively, a nova is bright
in optical light during the early evolutionary phases and, while it
declines in optical, it becomes brighter in the ultraviolet (UV) bands.
As long as the central nuclear burning rate remains stable, $L_{\rm bol}$
will not change while the peak of the spectral energy
distribution gradually shifts to higher energies. With the continued
outflow, the density of the outer ejecta goes down, becoming
transparent. Eventually, the photospheric radius shrinks to near the
surface of the white dwarf, and the observable spectrum then peaks in
the soft X-rays \citep[e.g.,][]{st08}. Most of the soft X-rays are
absorbed in the interstellar medium with typical values of neutral
hydrogen column densities several $10^{21}$\,cm$^{-2}$. The spectrum
observed from Earth is known as the class of Super Soft Sources (SSS).\\

A nova is called {\em recurrent} if more than one outburst has been observed,
obviously leading to a bias of novae with recurrence time scales
$<100$ years. Since the white dwarf is not destroyed during the outburst
and the binary orbit not significantly altered, all novae should be
recurrent. Recurrence time scales depend on the accretion rate (that
drives how quickly fuel can be accumulated) and the WD mass (that
drives the ignition conditions being easy or difficult to achieve).
For example, a higher WD mass allows ignition already at a lower amount
of accreted mass.\\

V3890\,Sgr is a recurrent ($\sim 30$ years) nova in a symbiotic binary
system and has a famous sibling of the same type, RS\,Oph ($\sim 20$
years, see, e.g. \citealt{nessrsoph}). The ejecta from the nova outburst
run into the dense wind of the evolved companion star where they dissipate
some of their kinetic energy in the form of shocks which is then released
as radiation in the X-ray band consisting of bremsstrahlung continuum
and characteristic collisionally excited emission lines that originate
from collisionally ionised plasma.\\

The distance to V3890\,Sgr can be assumed to be of order 9\,kpc
\citep{v3890sgrdist}. Depending on the method, smaller values
have been derived, e.g., 4.5\,kpc from equivalent widths of the interstellar
lines or 4.3\,kpc from Gaia DR2, however, \cite{v3890sgrdist} argue
that it is not helpful that the Gaia EDR3 parallax is much different from
the DR2 value. Also, \cite{schaefer_gaia} does not consider the Gaia parallax
sufficiently reliable to include V3890\,Sgr in their catalogue.
\cite{v3890sgrdist} derive the maximum distance set by the Roche Lobe
radius derived from the results of their analysis of orbital variability
and spectroscopic orbits. We caution, however, that this assumes Roche Lobe
overflow while for RS\,Oph (for which the Gaia
parallax is also not reliable) the distance of 1.6\,kpc has become the
canonical and unquestioned value, even though \cite{schaefer2009}
argues that at this distance, the companion star underfills its Roche Lobe.
The assumption of a Roche-Lobe filling companion would require a
distance to RS\,Oph of 7.3\,kpc. In the subsequent work, we assume
9\,kpc distance to V3890\,Sgr.\\

The orbital period has also been determined by \cite{v3890sgrdist}
$747.6$\,days from both, photometric and spectroscopic data.\\

The 3rd recorded outburst of V3890\,Sgr was discovered by A. Pereira on
2019 August 27.87 which is defined as the reference time $t_0$ during this
work. The optical peak may have occurred on or around August 27.75
\citep{strader2019}, and any day after $t_0$ given here can be converted
to days after optical maximum by adding 2.88 hours.\\

During the 2019 outburst of V3890\,Sgr
extensive observing campaigns in all wavelength bands have been performed.
X-ray monitoring observations of V3890\,Sgr with \swift, \nicer, and \asat\
with low temporal and spectral resolution but a large baseline allow studies
of the long-term evolution. These datasets are described in detail by
\cite{page20} and \cite{singh20}. A deep \chandra\ observation of the
early shock emission was taken on day 6 and is described by \cite{orio2020}.
We focus here on a deep \xmm\ observation of V3890\,Sgr taken on 2019
September 15 that contains both, shock emission and SSS emission.

\begin{figure*}[!ht]
\resizebox{\hsize}{!}{\includegraphics{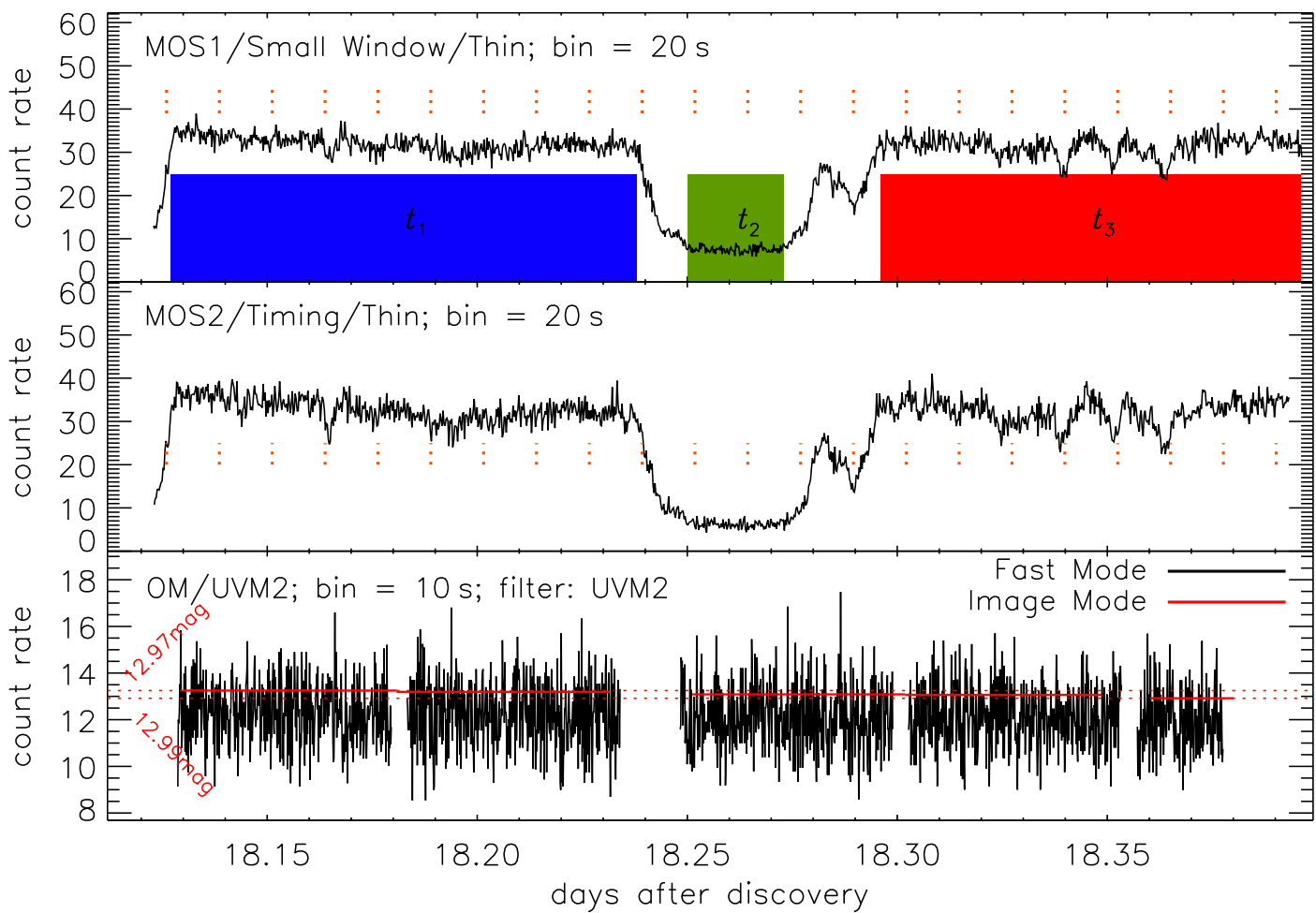}}%multilc
\caption{\label{fig:xmmlc}\xmm\ X-ray and UV light curves (units counts
per second) extracted from MOS1 (Small Window, 200-850\,eV range),
MOS2 (Timing Mode, 200-850\,eV range), and
the Optical Monitor (Image+Fast mode with the UVM2 filter:
$\sim 1620-2620$\,\AA\ range). In the top panel, shaded areas
mark three light curve segments analysed separately in 
\S\ref{sect:analysis:timing}. 22 short orange dotted vertical
lines are spaced by 18.1 minutes (centered in the secondary
dip within egress from the deep dip), corresponding to the
frequency of 0.92\,mHz discussed in \S\ref{sect:analysis:timing}.
Of these 22 marks, 8 (36\%) correspond to an anomaly in the light curve
(deep dip in-/egress or small dip) while the duty cycle increases
to 50\% (5/10) after the dip leading to a formally statistical detection.
In the bottom panel, the high time resolution OM Fast mode light curve
is shown in black. Three OM Image mode exposures were taken simultaneously
with each OM Fast mode segment, and the average Image mode count rates
are shown with the horizontal orange lines covering the start/stop times.
The red horizontal lines mark the magnitude range 12.97 to 12.99 within
which the sources gradually fades from start to the end of the observation.
This magnitude range is $\sim 5$ mag brighter than during quiescence
(see text). See \S\ref{sect:descr:lc} for discussion.
}
\end{figure*}

\section{Observations}
\label{sect:obs}

Before turning our attention to the 2019 observations of V3890\,Sgr,
we had a quick look at a pre-outburst 56.7ks \xmm\ observation
taken 2010 April 8 (ObsID 0604980101) but only found a $0.2 - 12$\,keV
X-ray upper limit of $5.1\times 10^{-15}$\,erg\,cm$^{-2}$\,s$^{-1}$
using the XMM Upper Limit Server
\footnote{\url{http://www.cosmos.esa.int/web/xmm-newton/uls}}.
The OM was operated in Image mode with multiple exposures in filters U, UVW1,
and UVM2 yielding magnitudes of 16.6, 16.9, and 18.4, respectively. The UVM2
magnitude during the new outburst observation is 13\,mag (see below),
thus 5 magnitude brighter than during quiescence. A linear decline rate
in mag would bring the UVM2 magnitude to 18.4 mag on day 72, however the
\swift/UVOT data show a non-linear decline behaviour, see 
\cite{page20}.\\

Since early hard shock interactions between the nova ejecta and the
dense stellar wind of the evolved companion star had been expected,
\swift\ monitoring started within hours after discovery with
$\sim 2$ observations per day. The campaign led to the surprisingly
early discovery of SSS emission 8-9 days after $t_0$ \citep{sw_sss}.
For RS\,Oph, first signs of SSS were only seen around day 26-29 after
outburst \citep{osborne11}.\\
As soon as the start of the SSS phase had been confirmed, an \xmm\
observation was triggered, however, it could not be performed before
September 10, owing to visibility
limitations. Further constraints led to the final observation being
taken September 15 (day 18.12 after $t_0$). Details of the exposures of
all five instruments are given in Table~\ref{tab:obs}. In anticipation
of extremely bright soft emission as seen in RS\,Oph \citep{nessrsoph},
the EPIC-pn instrument was operated in the very fast burst mode that
has a low duty cycle to better handle extremely bright sources. V3890\,Sgr
was in the end fainter than RS\,Oph, but a serious drawback of the burst
mode is that it is not sensitive below 0.7\,keV where the SSS emission
has its peak, and the pn data are thus not used in this work. The MOS1
camera was operated in Small Window mode in order to obtain an image
around the source which is important to identify any field sources that
might contaminate a Timing Mode observation. MOS2 was
operated in Timing Mode to minimise pile up at the expense of spatial
resolution. The RGS is the prime instrument to obtain a
high-resolution grating spectrum, and the Optical Monitor (OM) was operated
in Image+Fast mode in the UVM2 filter; an OM filter with higher throughput
may have led to coincidence losses. Standard extraction methods were
employed with SAS version 17.0.0 to extract images, light curves, and
spectra. The MOS1 extraction region was placed at pixel coordinates
(27044,27615) with a radius of 300 pixels, while the background was
extracted from (28068,26527) with radius 320 pixels. The MOS2 products
were extracted from the DETX pixel range 300-317.\\

A second observation was foreseen to be taken later during the SSS phase,
however, the nova turned off much sooner than expected. Therefore, only a single
X-ray grating observation was performed during the SSS phase and is presented
here.\\

\begin{table*}
\begin{flushleft}
\renewcommand{\arraystretch}{1.1}
\caption{\label{tab:obs}Journal of exposures taken with the \xmm\ observation ID 0821560201}
\begin{tabular}{llllll}
\hline
Instrument & Mode & Filter & UT start & UT stop & Ex. time (s)\\
pn & Timing/Burst & Thin & 2019-09-15@00:16:46 &2019-09-15@06:25:53 & 21677\\
MOS1 & Small Window & Thin & 2019-09-14@23:48:06 &2019-09-15@06:25:33&23667\\
MOS2 & Timing & Thin & 2019-09-14@23:48:22 &2019-09-15@06:21:24 &23402\\
RGS1 & standard & & 2019-09-14@23:47:02 &2019-09-15@06:26:48&23884\\
RGS2 & standard & & 2019-09-14@23:47:07 &2019-09-15@06:26:48 &23879\\
OM &  FAST & UVM2 &2019-09-14@23:58:10  &  2019-09-15@01:11:20 & 4390\\
OM &  FAST & UVM2 &2019-09-15@01:16:36  &  2019-09-15@02:29:46 & 4390\\
OM &  FAST & UVM2 &2019-09-15@02:50:04  &  2019-09-15@04:03:14 & 4390\\
OM &  FAST & UVM2 &2019-09-15@04:08:30  &  2019-09-15@05:21:40 & 4390\\
OM &  FAST & UVM2 &2019-09-15@05:26:57  &  2019-09-15@05:56:27 & 1770\\
\hline
\end{tabular}
\renewcommand{\arraystretch}{1}
\end{flushleft}
\end{table*}

The combination of \swift, \asat, \nicer, and \xmm\ observations taken during
the SSS phase is illustrated in
Fig~\ref{fig:mmlc}. The UV brightness experienced a slow gradual
decline, consistently seen in \swift/UVOT and \xmm/OM. Both \swift/XRT
and \asat/SXT show a highly variable bright episode between days 16
and 16.9, a faint episode without variability around days 16.9 to
17.9, and a brighter, less variable episode starting on day 17.9
until the end of the \asat\ campaign on day 19.5.
The \xmm\ observation was taken during the start of the second bright
episode, fortunately shortly after the end of the faint phase. It started
with a steep increase in brightness, consistent with \swift\ 
and \asat\ data and contained another dip between days 18.2-18.3 which is
not resolved in the \swift\ and \asat\ data.\\

\section{Description of the Data}
\label{sect:descr}

The most robust information can be deduced from the data, without any
bias from model assumptions. We thus present first the data in this
section with qualitative results while quantitative results can be found
in \S\ref{sect:analysis}.

\subsection{UV and X-ray Light Curves}
\label{sect:descr:lc}

\begin{figure*}[!ht]
\resizebox{\hsize}{!}{\includegraphics{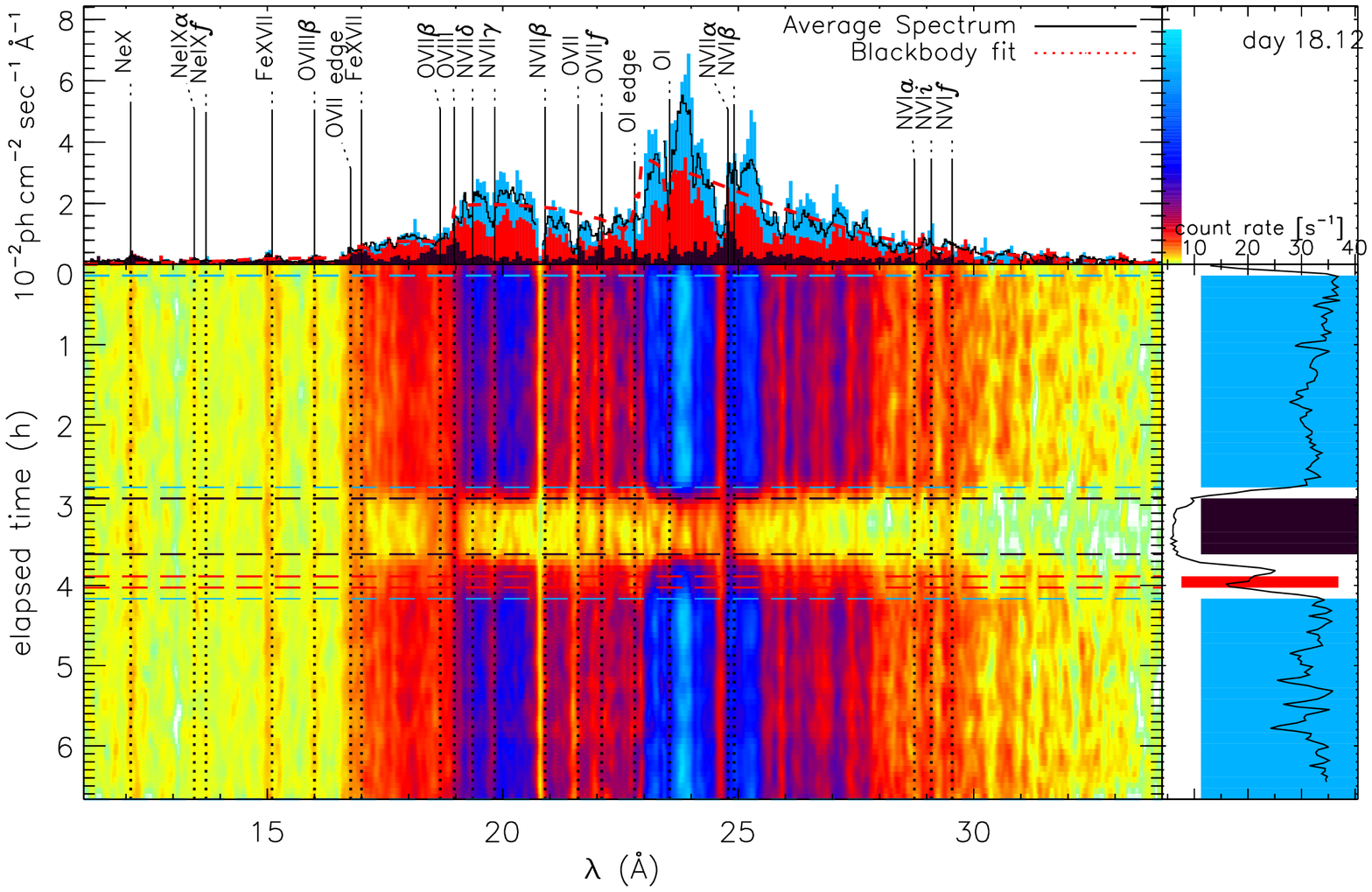}}%smap
        \caption{\label{fig:smap}Spectral Time map based on 48 RGS spectra
extracted from adjacent 500-s time intervals. The main panel shows the
spectra with wavelength across, time down, and flux colour coded following
the non-linear colours shown as bar in the top right panel along the vertical
flux axis. The shadings in the bottom right panel indicate time intervals
from which the spectra shown in the same colour shadings in the top left
panel have been extracted. In the top left panel, also the total spectrum
is shown, with a black solid line, and simple blackbody fit with the red
dotted line. See \S\ref{sect:descr:smap} for details.}
\end{figure*}

The X-ray and UV light curves extracted from EPIC/MOS and OM are shown
in detail in Fig.~\ref{fig:xmmlc}. The initial steep rise and the deep dip
are the most striking features in the X-ray light curves. The initial rise
indicates the launch of the second SSS bright phase after a fainter interval
of one day duration (Fig.~\ref{fig:mmlc}, \citealt{singh20}).
The deep dip lasted
$\sim 1$ hour with steep ingress and egress and roughly constant emission in
between. It was followed by a smaller dip within the rise back to the pre-dip
count rate. Further smaller dips can be seen, at least one before and three
after the main dip. The OM light curve shows a slow continuous declining trend from
mag 12.97 and 12.99. The decline is consistent with the \swift/UVOT data
(see Fig.~\ref{fig:mmlc}) yielding a decline rate of 0.1\,mag per day. The UV
light curve was unaffected by the events that caused the X-ray variability
which could mean that UV and X-ray emission come from different regions. If
the X-ray dip was caused by an occulting body that moved in front of the
regions emitting both UV {\em and} X-ray emission, it would consist of material
that is optically thick only to X-rays. For example, as argued by
\cite{nessusco}, neutral hydrogen is transparent to UV with energies
$<13.5$\,eV while X-rays suffer photo-electric absorption depending on
the column density.\\

\subsection{High-resolution Grating Spectra}
\label{sect:descr:grating}

Common spectral analysis approaches of low-resolution CCD spectra are based
on fitting a spectral model and drawing conclusions from the model.
Meanwhile, high-resolution grating spectra allow drawing conclusions directly
from the data because individual emission and absorption lines can be resolved.\\

\subsubsection{Overview and Spectral Evolution}
\label{sect:descr:smap}

In Fig.~\ref{fig:smap}, we illustrate
the time evolution of the RGS spectrum. We have produced 48 RGS events files
by filtering on adjacent time intervals of 500s duration and extracted
corresponding RGS spectra which are aligned as a time evolution map in the
main panel visualising the spectral evolution. In the
right panel, the light curve is shown along the vertical time axis with
count rate across and shadings marking time intervals from which a
bright (light blue) and a dip spectrum (dark purple) have been extracted.
The time stamp ranges are given for reproduction in Mission Time units
(seconds since reference MJD
50814=1998-01-01T00:00:00 TT) are $(6.8490264-6.8490537)\times 10^8$\,s for the
dip spectrum (2.7\,ks) and $(6.8489253-6.8490220)\times 10^8$\,s and
$(6.8490711-6.8491643)\times 10^8$\,s for the bright spectra before
and after the deep dip, respectively, with the summed exposure time of
19\,ks. These ranges exclude the early rise and the dip. We have also extracted
RGS spectra during the three smaller dips after the deep dip, combining
data from the time stamp ranges $(6.8491072-6.849111)\times 10^8$s,
$(6.8491183-6.8491218)\times 10^8$s, and
$(6.8491287-6.8491318)\times 10^8$s. Finally, we have extracted a separate
spectrum during the secondary dip within the egress part of the
deep dip, time stamp range $(6.8490613-6.8490688)\times 10^8$s.

A blackbody model was fitted to the combined bright spectrum in
an attempt to describe the overall spectral shape.
To account for photoelectric absorption by the cold interstellar
medium (ISM), we use the {\tt tbabs} model by \cite{wilms00}. The dominant
ISM absorption is due to neutral hydrogen at energies above
13.6\,eV ($\lambda<911$\,\AA), with exponentionally increasing transmission
towards higher energies, thus affecting long wavelengths more than shorter
wavelengths. The only model parameter is the neutral hydrogen
column density $N_{\rm H}$, and we found a best fitting value of
$N_{\rm H}=5.4\times 10^{21}$\,cm$^{-2}$.
Photoelectric absorption by other elements are included in the
{\tt tbabs} model with default cosmic abundances.
In order to best reproduce the O\,{\sc i} absorption edge at 22.8\,\AA,
we have reduced the abundance of oxygen by 30\%. Since the {\tt tbabs}
model only accounts for neutral absorption, in order to reproduce the
absorption edges of highly ionised N\,{\sc vi} and N\,{\sc vii},
two artificial absorption edges were included at 16.75\,\AA\ and
18.79\AA, with the column densities of $2.2\times 10^{17}$\,cm$^{-2}$
and $1.1\times 10^{17}$\,cm$^{-2}$, respectively, after manual adjustment.
For the parameter $T_{\rm eff}$, we adopted a value of
$6.9\times 10^5$\,K which roughly reproduces the overall spectral shape.
Note that the assumed depths of edges and O\,{\sc i} abundance have a
strong effect on the best-fitting $T_{\rm eff}$ and $N_{\rm H}$
parameters as already noted by \cite{page20}. We do not invest efforts
into resolving the degeneracy of these model parameters as the
(oversimplified) assumptions of a blackbody model are unphysical given
the complex circumstances.\\

\begin{figure*}[!ht]
\resizebox{\hsize}{!}{\includegraphics{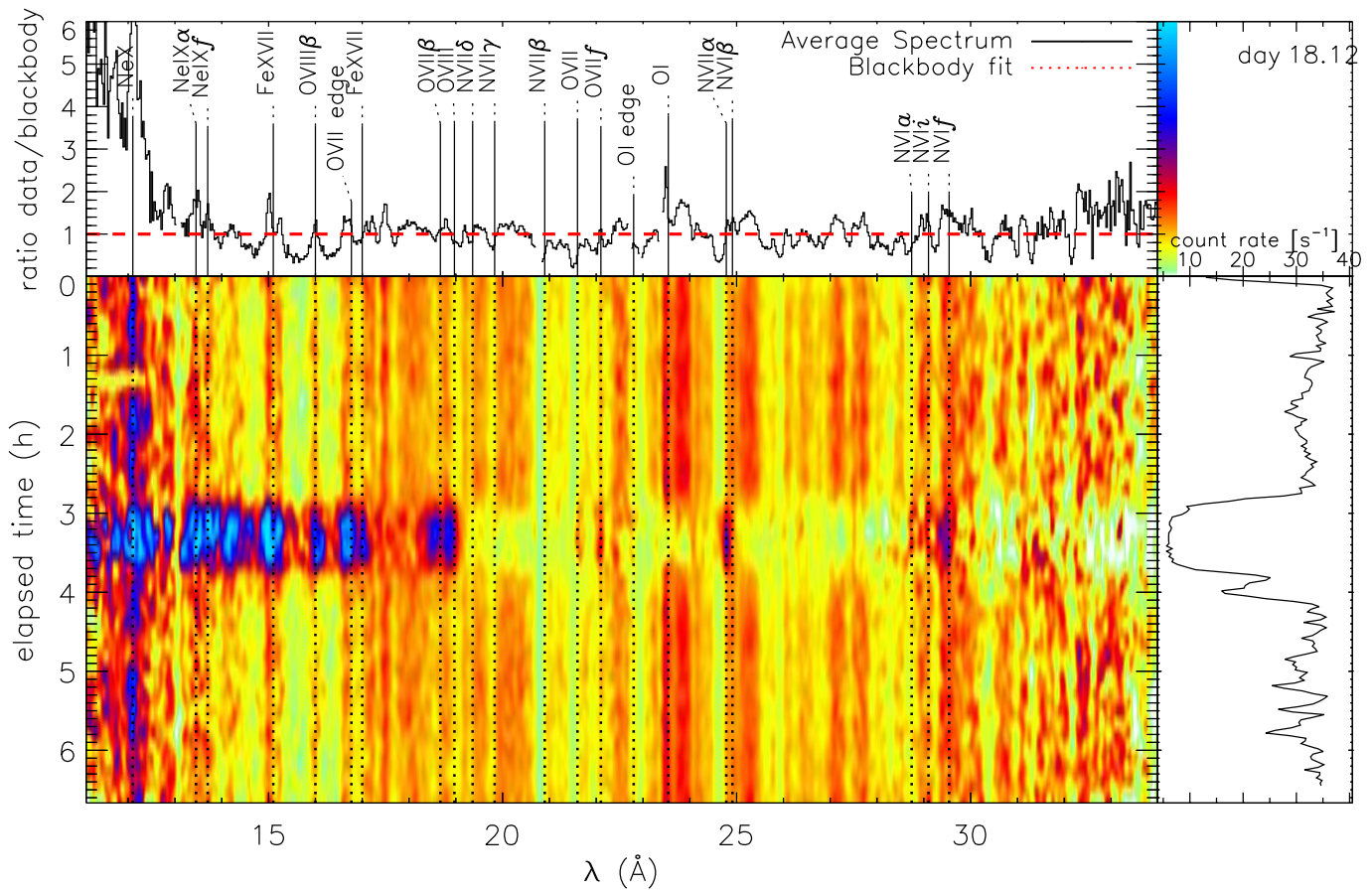}}%smap_bbnorm
        \caption{\label{fig:smap_bbnorm}Same as Fig.~\ref{fig:smap} where each
of the 48 spectra was divided by a blackbody curve with the same parameters
as the red dotted line in the top panel of Fig.~~\ref{fig:smap}, only
renormalised to the brightness of the respective spectrum. Wavelengths at
which the ratio is below one indicate absorption lines, e.g., N\,{\sc vii}
at 24.8\,\AA\ whereas wavelengths with ratios greater than 1 indicate
emission lines, e.g., the same N\,{\sc vii} or the N\,{\sc vi} forbidden
line at 29.1\,\AA\ during the dip. See \S\ref{sect:descr:smap} for details.
}
\end{figure*}

The blackbody fitting is not intended to derive physical quantities but
to identify trends, similar to \cite{page20} who studied the long
evolutionary trends, e.g., of $T_{\rm eff}$. In
Fig.~\ref{fig:smap_bbnorm} we show the evolution of the ratio of observed
spectrum to blackbody reference curve. For each of the 48 sub-spectra,
the same blackbody model was used as for the total spectrum except for the
normalisation that was adapted to the brightness evolution. In the map, red and blue
colours thus indicate excess above the blackbody line while any light
colours indicate reduced emission such as absorption lines. The vertical
dashed lines guide the eye demonstrating that absorption lines are blue
shifted, e.g., N\,{\sc vii} at 24.8\,\AA\ or O\,{\sc vii} at 21.6\,\AA.
Meanwhile, emission lines appearing strong relative to the blackbody
continuum during the dip are seen at rest wavelengths.\\

\subsubsection{Continued Shock Emission}
\label{sect:descr:shock}

\begin{figure}[!ht]
\resizebox{\hsize}{!}{\includegraphics{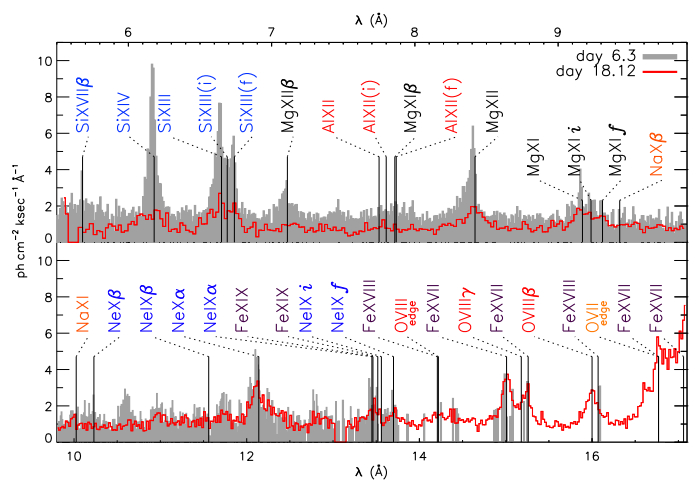}}%cmpv3890sgr.eps
\caption{\label{fig:shock}Comparison of the pre-SSS \chandra/HETG
spectrum taken on day 6.3 \citep[ObsID 22845;][]{orio2020} and the
\xmm/RGS spectrum taken on day 18.12. The line labels are placed
at rest wavelengths using different colours for different elements
for better distinction. See \S\ref{sect:descr:shock} for discussion.
}
\end{figure}

The early shock emission before the SSS phase started was observed
with \chandra\ on day 6.3 and has been analysed by \cite{orio2020}.
In Fig.~\ref{fig:shock} we show the hard part of the \xmm/RGS spectrum
compared to the earlier \chandra/HETGS spectrum. One can see that
the Si, and Mg\,{\sc xii} lines have faded while the (lower-temperature)
Mg\,{\sc xi} and Ne lines are observed at about the same strengths;
possibly also the (yet cooler) O\,{\sc viii} lines at 16 and 19\,\AA\
appear equally strong, although the \chandra/HETGS has a very low
sensitivity at those wavelengths.
This is clearly a cooling effect as the Si\,{\sc xiv}
(log\,$T_{\rm peak}=7.2$), Si\,{\sc xiii} (log\,$T_{\rm peak}=7.0$),
and Mg\,{\sc xii} (log\,$T_{\rm peak}=7.0$) lines are formed at higher
temperatures than the Mg\,{\sc xi} (log\,$T_{\rm peak}=6.8$),
Ne\,{\sc x} (log\,$T_{\rm peak}=6.8$), and Ne\,{\sc ix}
(log\,$T_{\rm peak}=6.6$) lines. The features at $\sim 13.5$\,\AA\
may be either Ne\,{\sc ix} or Fe\,{\sc xix} lines, the latter being
formed at log\,$T_{\rm peak}=6.8$ \citep{nebr}.
A slow reduction of temperature with time has been found from model
fits to the X-ray spectra from the \swift\ monitoring campaign
presented by \cite{page20}.\\
The asymmetric line profile of the Mg\,{\sc xii} line appears to have become
symmetric, however, close inspection of the line reveals that the profile
is consistent with the RGS line spread function and any asymmetry cannot
be resolved with the RGS.

\subsubsection{Identification of Absorption and Emission lines during the SSS phase}
\label{sect:descr:lines}

\begin{figure*}[!ht]
\resizebox{\hsize}{!}{\includegraphics{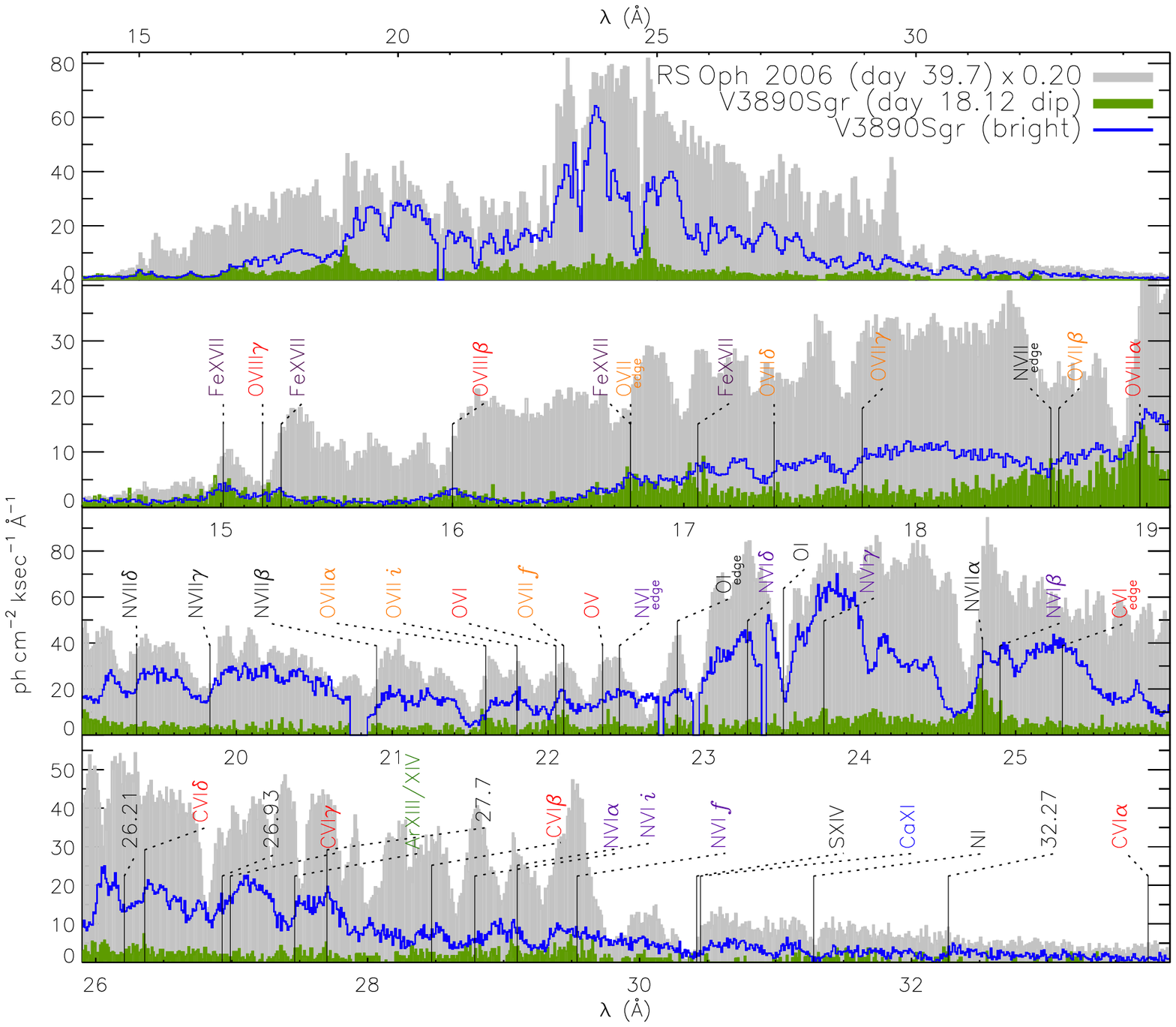}}%cmpv3890sgr_sss
        \caption{\label{fig:rgs}Comparison of RGS spectra of V3890\,Sgr
extracted during the dip (dark green shadings) and during time intervals
excluding the dip and the initial rise (blue line). The top panel shows
the full spectral range of interest while the panels below zoom into
more detail. The light grey
shading is the \chandra/LETGS spectrum of RS\,Oph taken on day
39.7 after outburst (Chandra ObsID 7296), scaled down by a factor 5.
See \S\ref{sect:descr:lines} for discussion. Labels with different colours
for different ions are included at their rest wavelengths.
}
\end{figure*}

In Fig.~\ref{fig:rgs}, we compare the bright spectrum (blue line)
and dip spectrum (dark green shadings) with RS\,Oph (light grey shadings)
during the early SSS phase \citep{nessrsoph}. Line labels of important
transitions are marked in order to validate which of them are detected.\\

At short wavelengths $\lambda<16.5$\,\AA, the bright spectrum and the dip
spectrum are identical indicating that the shock spectrum shown in
Fig.~\ref{fig:shock} was unaffected by the deep dip. This is a first
indication that the cause of the dip is to be searched for near
the surface of the white dwarf.\\

In the following we describe absorption and emission lines from selected
iso-electronic sequences. For the purpose of line identifications, we
ignore the blue shifts for the moment (thus line
labels and observed features do not exactly match) while we measure
them in \S\ref{sect:analysis:lprofiles}.\\

\noindent
{\bf N\,{\sc vii}}: The most prominent lines in SSS spectra of novae arise from
the H-like Ly series of nitrogen. The N\,{\sc vii} Ly\,$\alpha$ (1s-2p) line at
24.78\,\AA\ can be seen both as a deep absorption line during the bright
part and as an emission line in the dip spectrum. Also the Ly\,$\beta$ (1s-3p) line
(20.90\,\AA) is likely detected as an absorption line, although there is a
defect (chip gap) in the RGS1 at
20.81\,\AA\footnote{\url{https://xmm-tools.cosmos.esa.int/external/xmm\_user\_support/documentation/uhb/rgsmultipoint.html\#3861}},
and this wavelength range is not covered by the RGS2. The Ly\,$\gamma$ (19.82\,\AA, 1s-4p)
and Ly\,$\delta$ (19.36\,\AA, 1s-5p) lines are clearly seen in absorption, and also
the absorption edge at 18.6\,\AA\ is clearly visible.\\

\begin{figure}[!ht]
%\resizebox{\hsize}{!}{\includegraphics{o8_diff}\includegraphics{n7_diff}}
\resizebox{\hsize}{!}{\includegraphics{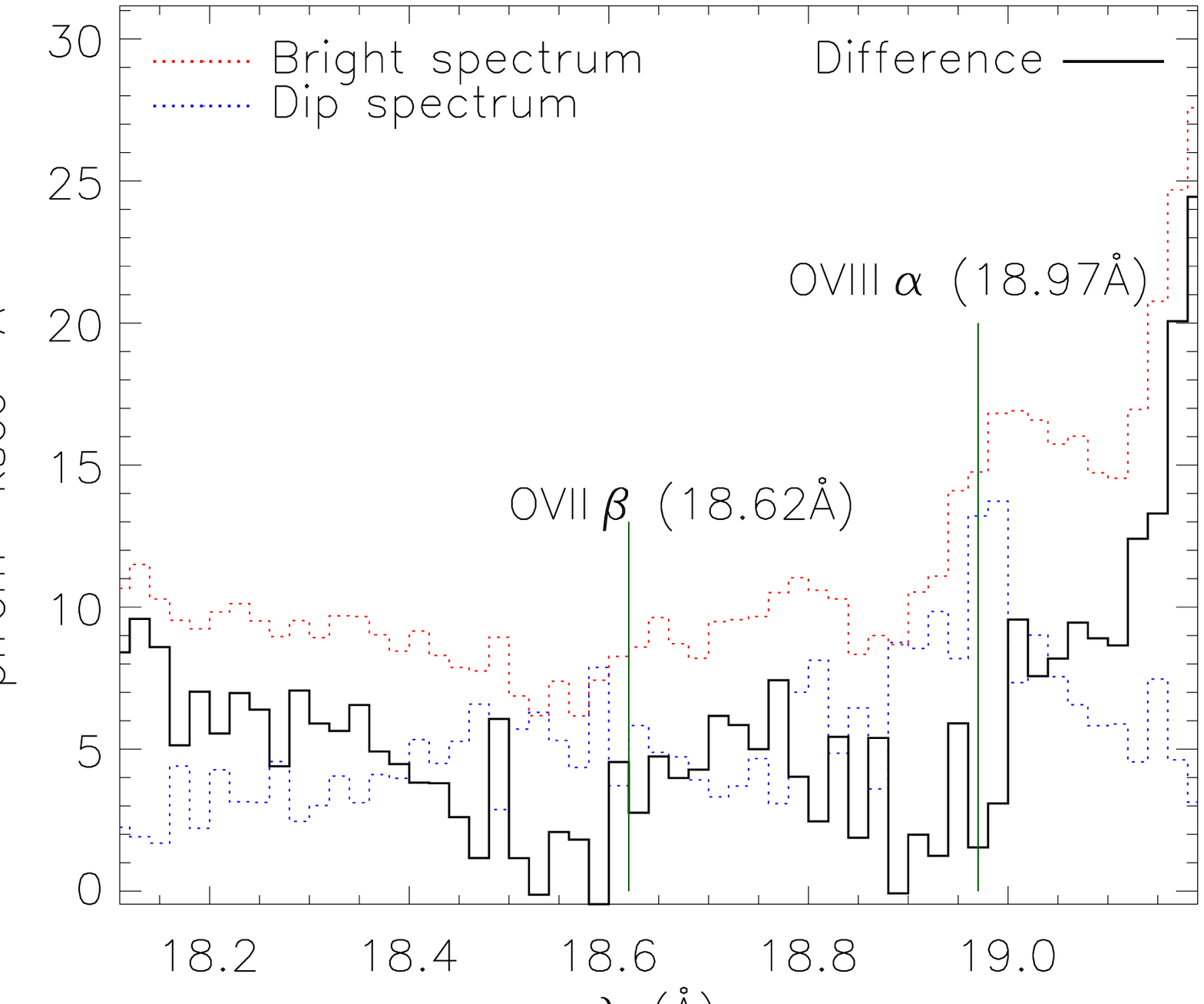}\includegraphics{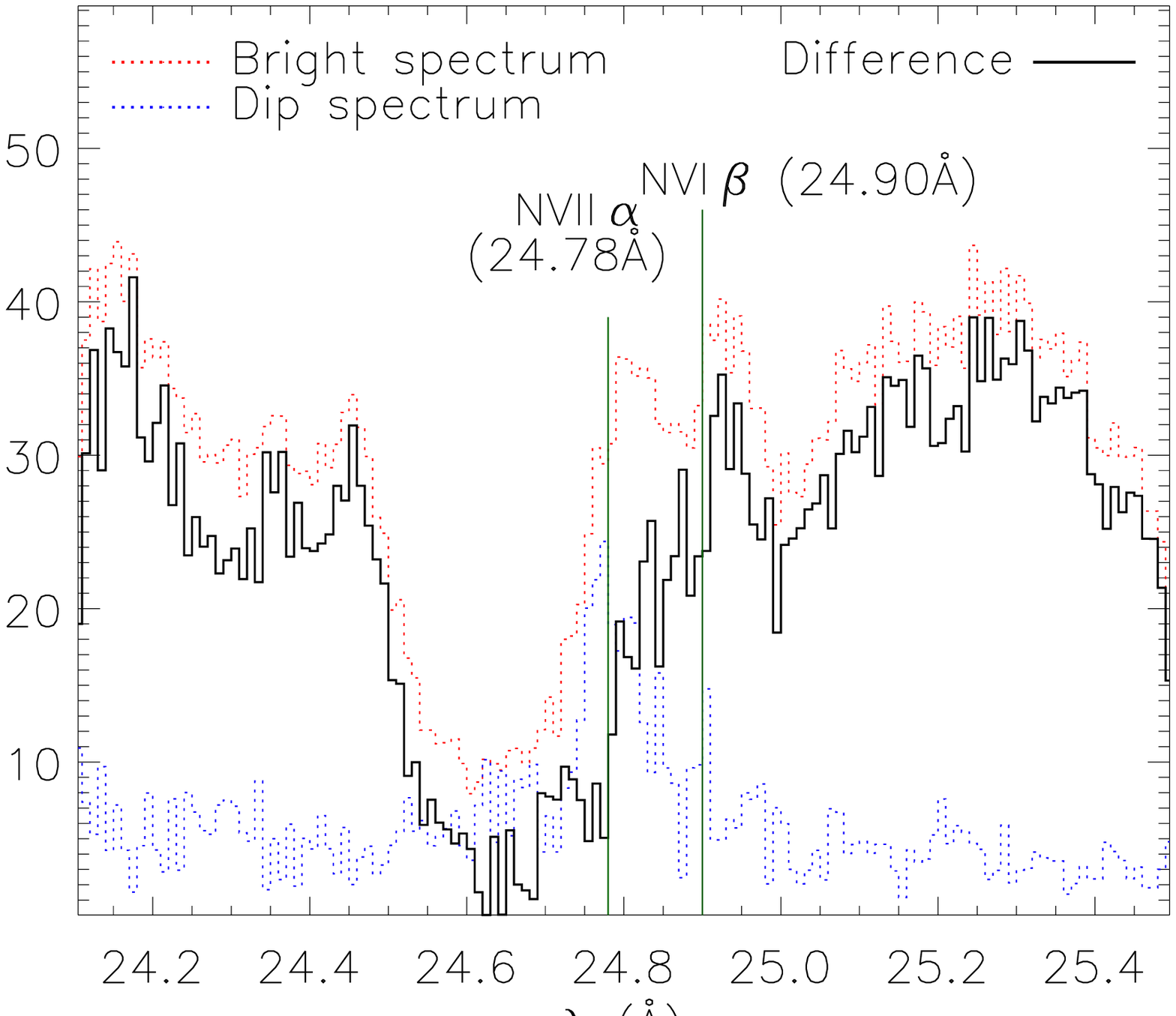}}
\caption{\label{fig:diff}The dip spectrum, shown with blue dotted lines,
most likely represents emission that is also present during the
bright spectrum (red dotted lines). It probably does not originate
from the SSS photosphere but further outside, likely the shock emission
site. As an illustration of how the pure SSS spectrum may look, the difference
between bright and dip spectrum is shown with black
solid lines for Oxygen (left) and the nitrogen lines (right), see
labels. See \S\ref{sect:descr:lines} for discussion.
}
\end{figure}

{\bf N\,{\sc vi}}: Meanwhile, the He-like nitrogen lines are much less
dominant. While in RS\,Oph, the N\,{\sc vi} 1s-2p line at 28.78\,\AA\
can clearly be seen in emission and absorption (also the intercombination
and forbidden lines in emission), the features in the spectrum of
V3890\,Sgr can not clearly be attributed to any of the N\,{\sc vi} lines.
Possibly (see discussion later), diffuse line emission and absorption
balance out. In fitting a blackbody continuum and looking
at the residuals (see Fig.~\ref{fig:smap_bbnorm}), an absorption feature
at roughly the same blue shift as for N\,{\sc vii} might possibly be
attributed to N\,{\sc vi}\,He\,$\alpha$\footnote{He\,$\alpha$ denotes
a 1s-2p transition in a He-like ion, He\,$\beta$ 1s-3p etc}, and possibly also the intercombination
and forbidden lines. Especially the forbidden line of
N\,{\sc vi}\,He\,$\alpha$ at 29.54\,\AA\ appears prominent during the dip.\\
The N\,{\sc vi}\,$\beta$ line at 24.9\,\AA\ blends with the
N\,{\sc vii}\,$\alpha$ line and could contribute to the broad absorption
feature. The emission line spectrum during the dip is dominated by
the N\,{\sc vii} line at 24.78\,\AA\ while there is a somewhat weaker
feature at 24.9\,\AA\ (best seen in Fig.~\ref{fig:n7}) that appears to
be N\,{\sc vi}\,He\,$\beta$. Also a weak N\,{\sc vi} absorption line
can be seen in the difference spectrum between bright and dip
spectra (see right panel of Fig.~\ref{fig:diff}).\\
At the wavelengths of N\,{\sc vi}\,$\gamma$ (23.77\,\AA) and
$\delta$ (23.28\,\AA), no convincing absorption feature is seen, as
well as no edge at the ionisation energy of N\,{\sc vi} to N\,{\sc vii}
at 22.46\,\AA.\\

\noindent
{\bf O\,{\sc viii}}: The H-like Ly series lines of oxygen are clearly
seen in RS\,Oph with the Ly\,$\alpha$ line at 18.97\,\AA, $\beta$ at 16\,\AA,
and possibly $\gamma$ at 16.77\,\AA. The SSS spectrum of V3890\,Sgr has
lower continuum at the wavelengths of the O\,{\sc viii} Ly
series lines, but the Ly\,$\alpha$ and $\beta$ lines can be seen in
emission in the dip spectrum. Especially for the O\,{\sc viii}\,Ly\,$\beta$
line, the emission in the dip and bright spectra are perfectly consistent
with each other (as also the Fe\,{\sc xvii} and Ne\,{\sc x} lines).
This indicates that the bright spectrum may be a composite
of the dip emission and the photospheric SSS emission. Assuming the dip
emission originates in the shock emission site (possibly a combination
of collisional and photo-ionisation/excitation), the pure SSS spectrum
can be reconstructed by subtracting the dip spectrum from the bright
spectrum. The result is shown for the wavelength region around the
O\,{\sc viii}\,Ly\,$\alpha$ line in the left panel of
Fig.~\ref{fig:diff}; the nearby O\,{\sc vii}\,He\,$\beta$ line at 18.62\,\AA\
is also included. While the bright spectrum shows little evidence of
an absorption line, the difference spectrum clearly shows both lines
as blue-shifted saturated absorption lines. Also the N\,{\sc vii} and
N\,{\sc vi} blend (right panel) appears saturated with the two lines
clearly separated as in the bright spectrum.\\

\noindent
{\bf O\,{\sc vii}}: The He-like O\,{\sc vii}\,He\,$\alpha$ line (21.6\,\AA) can
clearly be seen in Fig.~\ref{fig:rgs} in absorption and also in emission
during the dip spectrum. The He\,$\beta$ (18.62\,\AA) line is only seen as
an absorption line in the difference spectrum, left panel of
Fig.~\ref{fig:diff}. The He\,$\gamma$ and $\delta$ lines are clearly detected
as blue-shifted absorption lines with at best weak emission in the dip
spectrum. They are suspiciously broad and deep and one may wonder whether
they could be contaminated by other lines but no transitions of sufficiently
abundant elements with comparably high oscillator strengths are known
at these wavelengths.\\

\noindent
There is no clear evidence for carbon lines from the various transitions
marked in Fig.~\ref{fig:rgs}.\\

\begin{figure}[!ht]
\resizebox{\hsize}{!}{\includegraphics{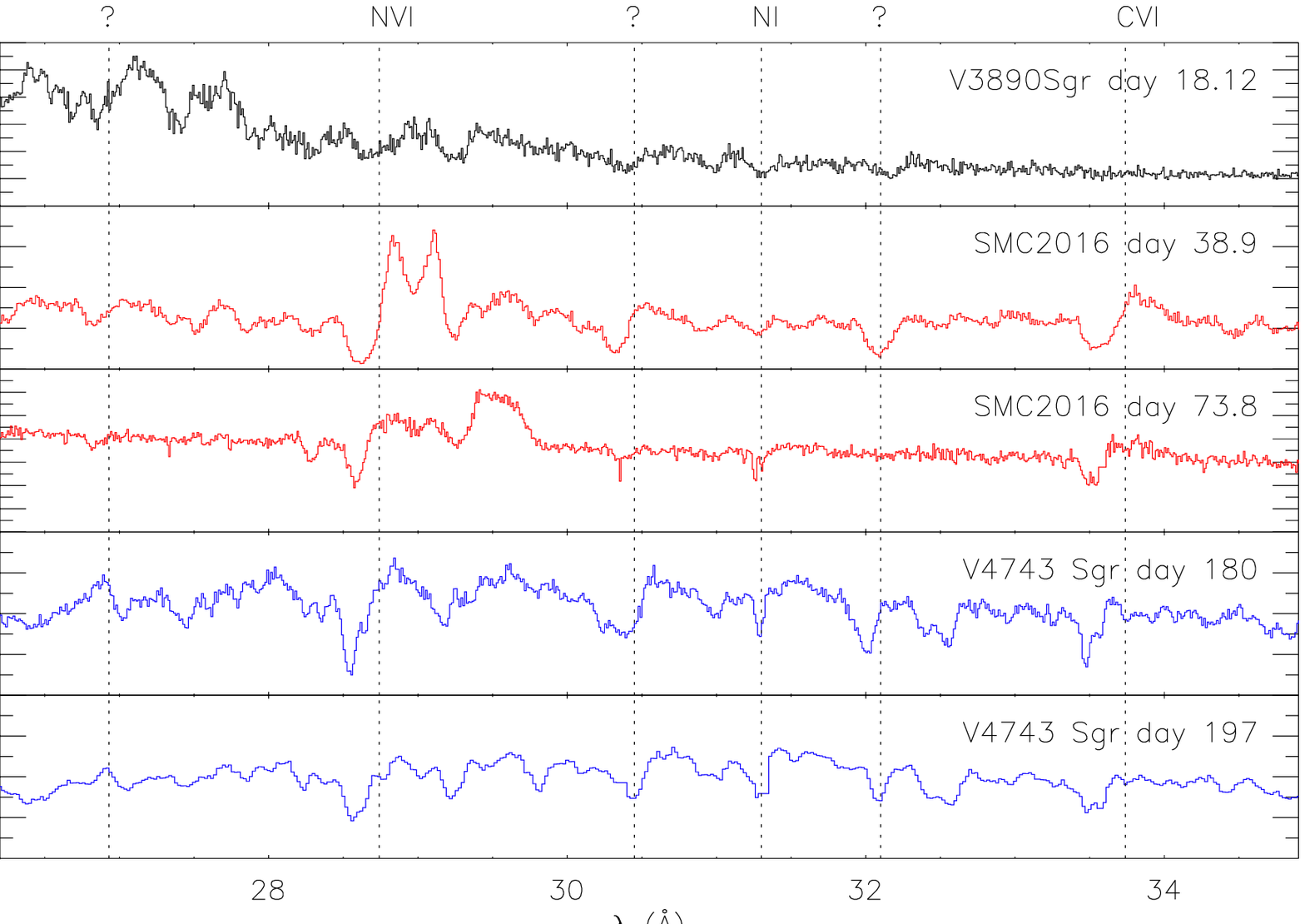}}
\caption{\label{fig:unknown}Comparison of spectral range between 26-35\,\AA\
of V3890\,Sgr with the novae SMC\,2016 on days 38.9 (\chandra\ Obs\-ID 19011) and
73.8 (\xmm\ ObsID 0794180201) and V4743\,Sgr on days 180
(\chandra\ Obs\-ID 3775) and 197 (\xmm\ ObsID 0127720501). Dotted
vertical lines mark wavelengths of absorption features to draw attention
to, with labels (only if unambiguously known) or question marks given in
the top. See \S\ref{sect:descr:lines} for discussion.
}
\end{figure}

\noindent
Other nova SSS spectra that have been studied in greater detail have shown
absorption features which could not easily be identified see, e.g., table~5 in
\cite{nessv2491}. We have searched for possible features around the
wavelengths listed as projected wavelengths (thus assuming they are blue
shifted by the same amount as known lines in the same spectra). In
Fig.~\ref{fig:unknown} we compare V3890\,Sgr with two other novae and
mark well known and less known absorption features.\\

A feature at 26.93\,\AA\ listed in table~5 of \cite{nessv2491} is seen
in V3890\,Sgr as possible blue-shifted narrow absorption line and in
V4743\,Sgr as emission line at rest wavelength (see also \citealt{nesscospar}). The Chianti atomic
database \citep{chianti} contains a C\,{\sc vi} 1s-4p line (26.9\,\AA)
and an Ar\,{\sc xv} 2p-3d line at 27.04\,\AA. Since there are no other
carbon lines detected, the Ar\,{\sc xv} identification appears more likely.
In emission, this line would be the strongest in the multiplet while another
Ar\,{\sc xv} 2p-3d transition at 26.5\,\AA\ has a higher oscillator strength
of $g_f=5.1$ while Chianti lists $g_f=1.6$ for the former transition. We do
not see an absorption feature around 26.5\,\AA\ and either this is another
unknown transition, or the oscillator strength in the Chianti database is
incorrect.\\

The N\,{\sc vi} 1s-2p line is only seen as a weak feature in V3890\,Sgr
(likely owing to overlapping weak SSS
emission and strong emission from the shock regions, see discussion above)
but as clear narrow blue-shifted absorption line in the other spectra,
where some emission component can also be seen in SMC\,2016 on day 38.9.\\

The feature at 30.42\,\AA\ from table~5 in \cite{nessv2491} can possibly
be seen in V3890\,Sgr but would not be identified as an absorption feature
without the knowledge of an absorption line being seen in other systems
as can be seen in the spectra below. The Chianti database lists as possible
candidates S\,{\sc xiv} 2p-3d at 30.30\,\AA, S\,{\sc xiv} 3s-3p at
30.43\,\AA\ or Ca\,{\sc xi} 2p-3d at 30.45\,\AA. The first transition has
a high oscillator strength of $g_f=4.6$, however, the line would then be
red-shifted, against the trend of other lines. The second listed transition
has a rather low oscillator strength of $g_f=0.46$, and stronger S\,{\sc xiv}
would need to be present. The Ca\,{\sc xi} 2p-3d transition is the strongest
in the multiplet and appears the most likely identification.\\

The N\,{\sc i} interstellar absorption line is always seen at 31.28\,\AA,
but the continuum in V3890\,Sgr is already quite weak.\\

\begin{figure}[!ht]
\resizebox{\hsize}{!}{\includegraphics{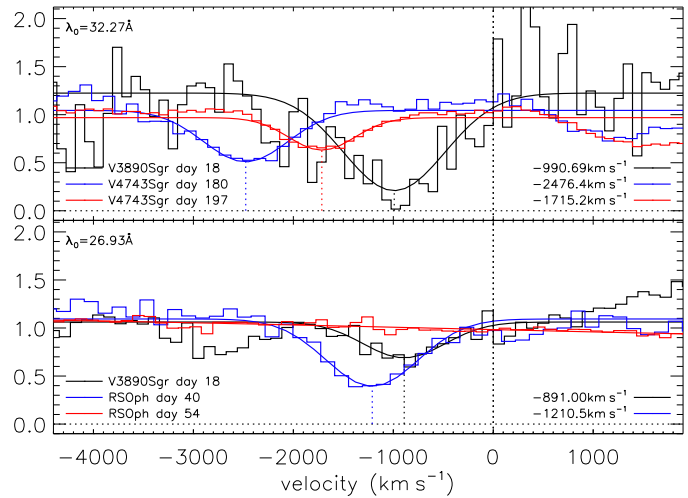}}%lprof_32}}
\caption{\label{fig:32}Comparison of line profiles of unidentified lines.
{\bf Top}:
Assumed rest wavelength 32.27\,\AA\ compared to V4743\,Sgr
on days 180 (\chandra\ ObsID 3775) and 197 (\xmm\ ObsID
0127720501).
{\bf Bottom}: Assumed rest wavelength 26.93\,\AA\
compared to RS\,Oph on days 40
(\chandra\ ObsID 7296) and 54 (\xmm\ ObsID 0410180301).
See \S\ref{sect:descr:lines} for discussion.
}
\end{figure}

A feature at $\sim 32$\,\AA\ was not listed in table~5 of \cite{nessv2491} but
was discussed in \cite{nesscospar} as being highly 'mobile': In
SMC\,2016, it disappeared between days 38.9 and 73.8 whereas in
V4743\,Sgr, it has shifted between days 180 and 197. This shift is shown
in more detail in the top panel of Fig.~\ref{fig:32} in velocity space.
The location of the absorption line
around 32\,\AA\ has changed by $\sim 500$\,km\,s$^{-1}$ within
only 2 weeks. No other absorption line has shifted by this amount
\citep{nesscospar}. The line is clearly detected in V3890\,Sgr at lower
blue shift, although the rest wavelength is obviously highly uncertain,
and with only one observation we do not know whether such a shift may
have occurred before or after this observation.
The Chianti database lists as possible transitions
S\,{\sc xiii} 1s-3p at 32.24\,\AA\ or S\,{\sc xiv} 2p-3d at 32.56\,\AA.
The S\,{\sc xiii} line with the highest oscillator strength is expected
at 34.85\,\AA\ ($g_f=5$ compared to $g_f=0.5$ for the 32.24-\AA\ line)
where we do not see an absorption line, although the continuum is already
very low. The S\,{\sc xiv} 2p-3d line at 32.56\,\AA\ has a value of $g_f=2.4$,
but another S\,{\sc xiv} 2p-3d at 30.77\,\AA\ is listed with $g_f=2.8$
which is not as clearly seen.\\

In the bottom panel of Fig.~\ref{fig:32}, the 26.93\,\AA\ region is shown
in comparison to RS\,Oph where an absorption line seen only on day 40
had disappeared 2 weeks later.
The rest wavelengths of the two lines are assumed here to yield the same
blue shift in V3890\,Sgr.

\subsubsection{The dip spectrum}
\label{sect:descr:dip}

\begin{figure}[!ht]
\resizebox{\hsize}{!}{\includegraphics{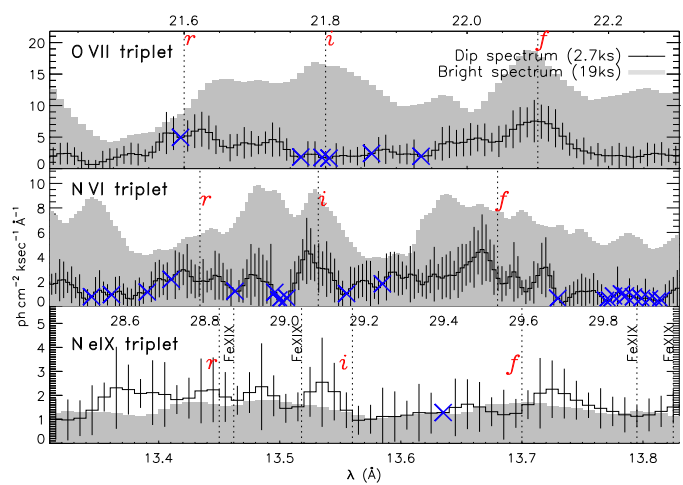}}%helike
\caption{\label{fig:he}He-like triplet lines of O\,{\sc vii} (top),
N\,{\sc vi} (middle), and Ne\,{\sc ix} (bottom) during the dip (black
histogram, 2.7\,ks) and during the bright time intervals (light
shadings, 19\,ks). The blue crosses mark bad
pixels with suspicious or invalid flux values. Dotted vertical lines
mark the resonance ($r$: 1s2p\,$^1$P$_1$)
intercombination ($i$: 1s2p\,$^3$P$_1$), and forbidden
($f$: 1s2s\,$^3$S$_1$) lines. In the bottom panel, potentially
blending lines of Fe\,{\sc xix} are also marked and labelled.
All spectra are smoothed with a Gaussian kernel of 0.03\,\AA, the
instrumental broadening, for better visualisation.
See \S\ref{sect:descr:dip} for discussion.
}
\end{figure}

\noindent
The dip spectrum is dominated by emission lines which can either be formed
by collisional or radiative interactions. In Fig.~\ref{fig:he}, the
dip spectrum is shown by zooming into the He-like triplet lines of
O\,{\sc vii}, N\,{\sc vi}, and Ne\,{\sc ix}. The He-like triplets consist
of a 1s\,$^1$S$_0$-2p\,$^1$P$_1$ resonance line, usually labelled with
$r$ and two forbidden lines: an
intercombination line 1s\,$^1$S$_0$-2p\,$^3$P$_1$ (thus forbidden spin
flip), labelled as $i$ and a forbidden line
1s\,$^1$S$_0$-2s\,$^3$S$_1$ (thus forbidden spin flip plus forbidden
$l=0$ transition), commonly labelled as $f$. The intercombination
line had been seen already in the 1960s in solar spectra as a result of the low
density in which the correspondingly low collision rates allow enough
time for the excited states of forbidden transitions to radiate back to
the ground. Meanwhile,
the forbidden line was not expected to be seen because of the even
longer radiative de-excitation time scales, yet a strong line was seen
at 22.1\,\AA\ leading \cite{gj69} to do careful calculations, ultimately
demonstrating that the densities were indeed low enough to allow the
forbidden line of O\,{\sc vii} to be formed and to escape.
The hugely different excitation and de-excitation probabilities of these
lines, formed by the same element in the same ionisation stage, offer a
large range of plasma diagnostics independent of elemental abundance
and of a plasma within a narrow temperature range. The various
combinations of ratios are powerful plasma diagnostics, namely:
%\vspace{-.6cm}
\begin{enumerate}
   \setlength{\itemsep}{0pt}%
   \setlength{\parskip}{0pt}% 
\item $R=f/i$ depends on the plasma density: at increasing density,
   the timescale for collisional excitations of 1s2s\,$^3$S$_1-$1s2p\,$^3$P$_1$ 
   (thus $f\rightarrow i$) becomes shorter than de-excitations
   1s2s\,$^3$S$_1-$1s2p\,$^1$P$_1$ ($f$ line to the ground),
   and the $i$ line increases at the expense of the $f$ line.
\item $R=f/i$ also depends on the intensity of the UV radiation field: the
    $f\rightarrow i$ can also be triggered by photons at the right
    energy which lies in the UV. Independent measurements of the
    UV intensity at the energy $E(i)-E(f)$, combined with measurements
    of the $R$ ratio allows calculation of the dilution factor and thus
    the distance between UV emission source and X-ray plasma.
\item $G=(f+i)/r$ depends on the plasma temperature in a collisional
    plasma: The collisional excitation probabilities of the $r$ line
    depend differently on temperature than $i$ and $f$ lines.
\item $G=(f+i)/r$ is an indicator to distinguish photo-ionised from
    collisionally ionised plasmas: In a photoionised plasma, dominated
    by radiative recombination processes, the forbidden lines $i$ and
    $f$ are favoured over the $r$ line, leading to a higher $G$ ratio
    whereas a collisional plasma yields $G\sim 1$.
\end{enumerate}

In light of these rich diagnostics possibilities, we have a closer
look at the He-like triplets of various ions that probe different
temperature regions.
The signal to noise is very low in the dip spectrum, owing to the short
duration of the dip, and determination of accurate line fluxes is beyond
reach. However, the detection or non-detection alone
of one or the other line can give us qualitative information:\\

The best case is O\,{\sc vii} and this is therefore shown in the top panel
of Fig.~\ref{fig:he}. 
The $r$ and the $f$ lines can clearly be seen while the $i$
line is not detected in the dip spectrum. Since all lines are broadened,
any measured flux depends on the number of spectral bins included and
would thus be highly uncertain. Nevertheless, the absence of the
$i$ and detection of $r$ and $f$ lines at about the same strength suggests we
are dealing with a low-density plasma dominated by collisional
interactions. Also the intensity of any UV radiation at the site of
O\,{\sc vii} emission must be low.\\
The presence of a few bad pixels around 21.8\,\AA\ may cast some doubt on
the non-detection of the $i$ line, but we argue that the $i$ line should also
be broadened as the $r$ and $f$ lines in which case we should see some
emission outside the bad pixels. This can be seen in the bright spectrum 
(which also contains a suspicious pixel) where a broad emission feature
can clearly be seen around the $i$ line which is in fact seen equally
strong as the $f$ line. The $r$ line can only be seen as an absorption line
in the bright spectrum while the emission line appears not as substantially
stronger during the bright phase as the $i$ line. All this evidence
suggests that during the dip phase we only see the emission originating
from the shocked plasma that is collisionally dominated while during
the bright phase we see additional photoexcited plasma.\\

The N\,{\sc vi} and Ne\,{\sc ix} lines probe cooler and hotter plasma,
respectively, and are thus also of interest. For N\,{\sc vi}, the only
line that appears clearly detected is the $i$ line and perhaps the $f$ line,
although both would be blue shifted by a few pixels ($\sim 500$\,km\,s$^{-1}$).
There is no emission feature that could be attributed to the $r$ line which
complicates the interpretation. Possibly the $r$ line is self-absorbed owing to
a geometry preferably scattering out of the line of sight; this would not
affect the $i$ and $f$ lines as they have lower radiative excitation
probabilities. The bright spectrum shows a similarly confusing picture,
and drawing any conclusions from the N\,{\sc vi} triplet is thus difficult.

The Ne\,{\sc ix} triplet can be blended with Fe\,{\sc xix}, see
\cite{nebr}, but the apparent emission features appear in random places and
may just be noise. Since the 13-\AA\ region contains no SSS emission from
the WD photosphere, the bright spectrum (of longer duration) should not
be different and shows none of the apparent emission features. The effect
of resonant scattering into the line of sight only during the dip spectrum
appears unlikely in this case as none of the spikes can be associated to
a resonance emission line. We thus consider the Ne\,{\sc ix} triplet not
detected.

\section{Analysis}
\label{sect:analysis}

In the previous section, all conclusions that can be drawn directly from
the data are presented. They are the most robust conclusions because
they only depend on the quality of the data and calibration. In this
section we derive more quantitative conclusions which have a higher
physical value but require models that depend on more
assumptions and on the quality of atomic data bases.

\subsection{Absorption Line Profiles}
\label{sect:analysis:lprofiles}

The absorption line profiles provide information about the opacity distribution
within the ejecta and thus the radial density distribution. It is important to
assess the degree of complexity of the density profiles in the various absorption
lines for the interpretation of global spectral models that usually assume some
homogeneous density distribution in an easy-to-handle geometry (e.g., spherical
symmetry).\\

We use the method described by \cite{ness09} and \cite{nessv2491} to determine
line shifts $\lambda_{\rm shift}$, width $\lambda_{\rm width}$, optical depths at line centre $\tau_{\rm c}$, and line column densities $N_{\rm X}$
with results given in Table~\ref{tab:lprof}. This approach assumes a Gaussian
optical depth profile $\tau(\lambda)$, and after fitting the parameters
to the observations, $N_{\rm X}$ is computed by integration over
$\tau(\lambda)$ using oscillator strengths extracted from the Chianti database
\citep{chianti}. The challenges when fitting
the line profiles are how to deal with overlapping (blending) lines and
how to account for the role of the dip spectrum that radiates emission
which certainly should also be present during the bright spectrum.\\

\begin{figure*}[!ht]
\resizebox{\hsize}{!}{\includegraphics{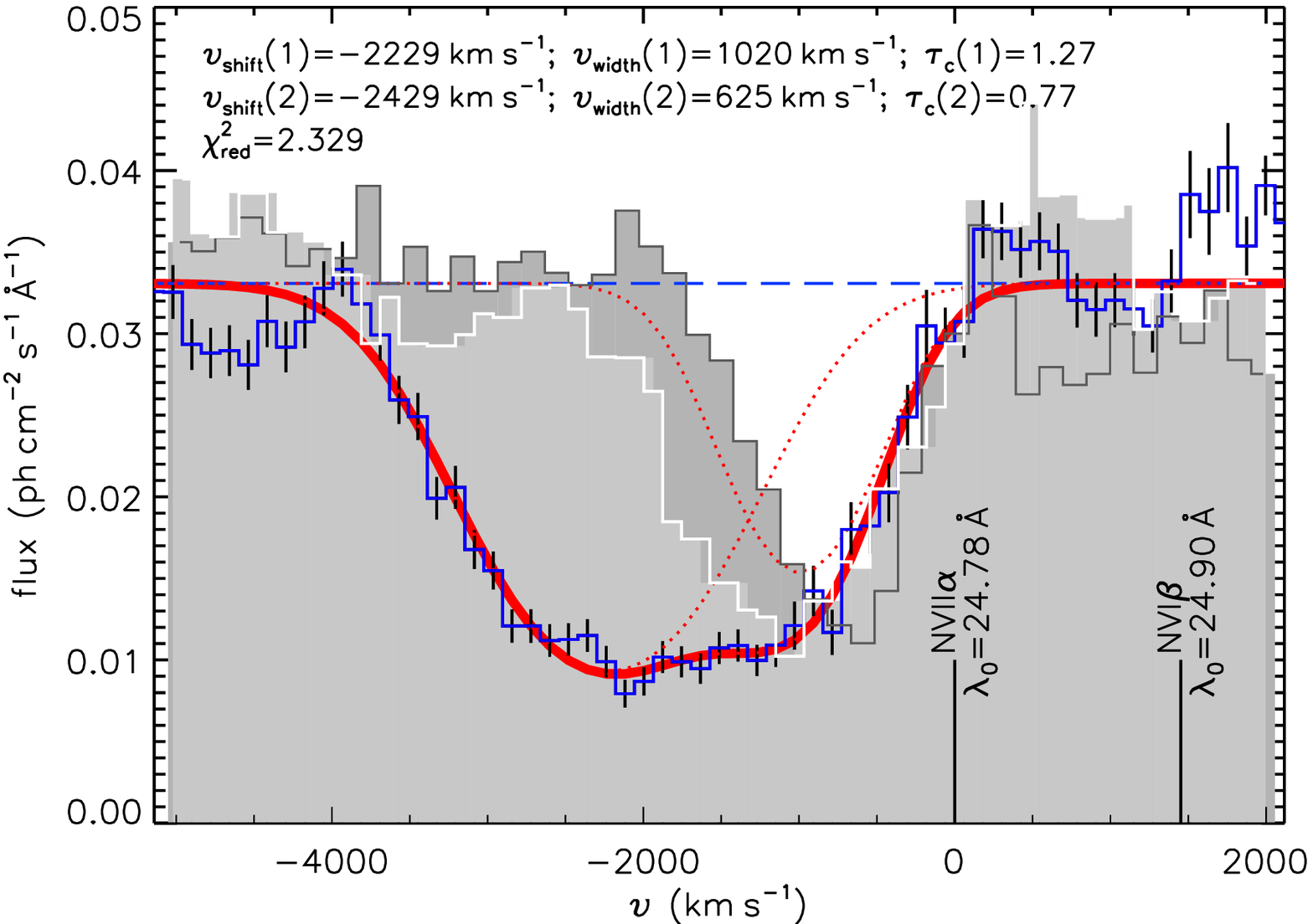}\includegraphics{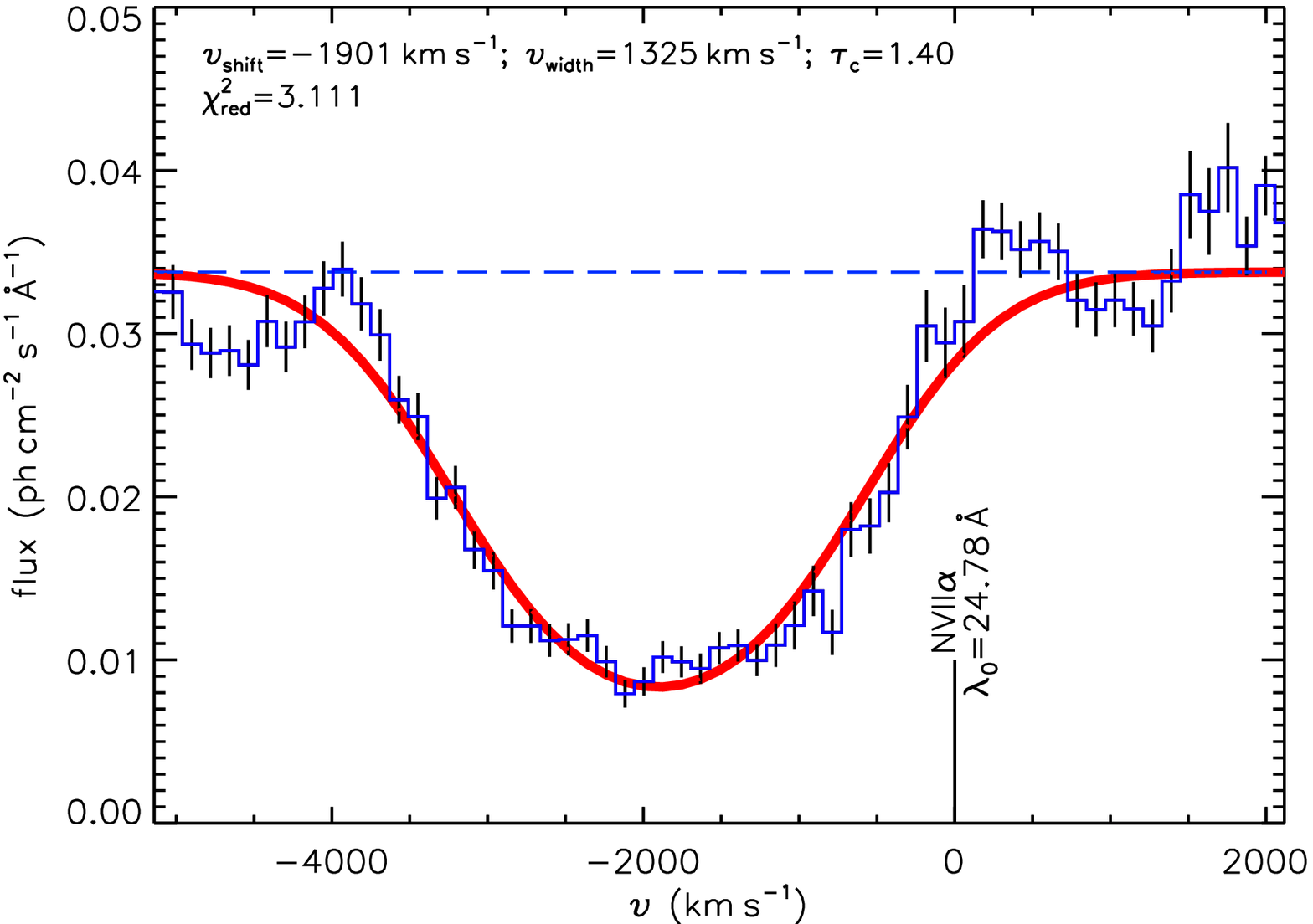}}%n76_rsoph}{n7}

\resizebox{\hsize}{!}{\includegraphics{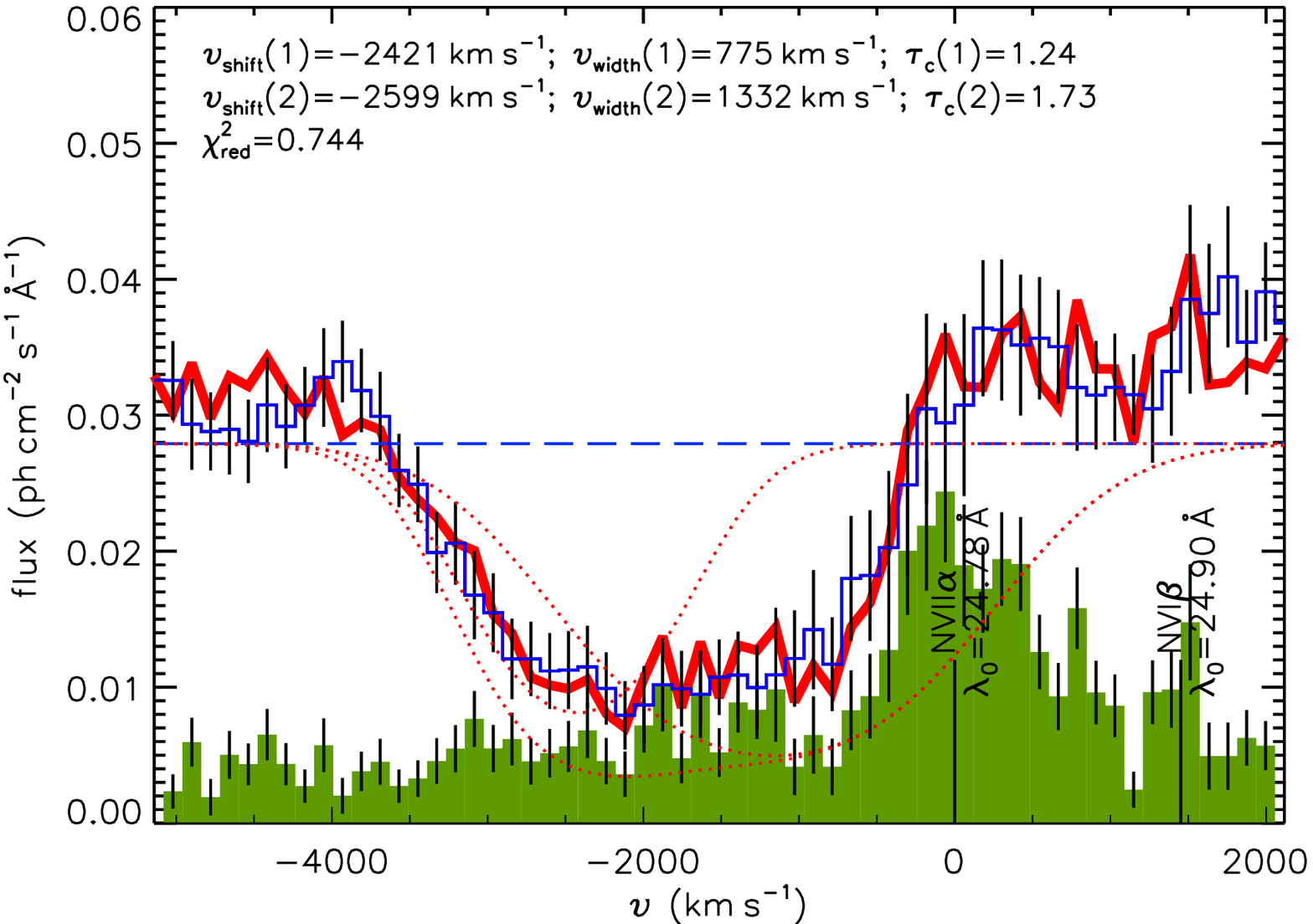}\includegraphics{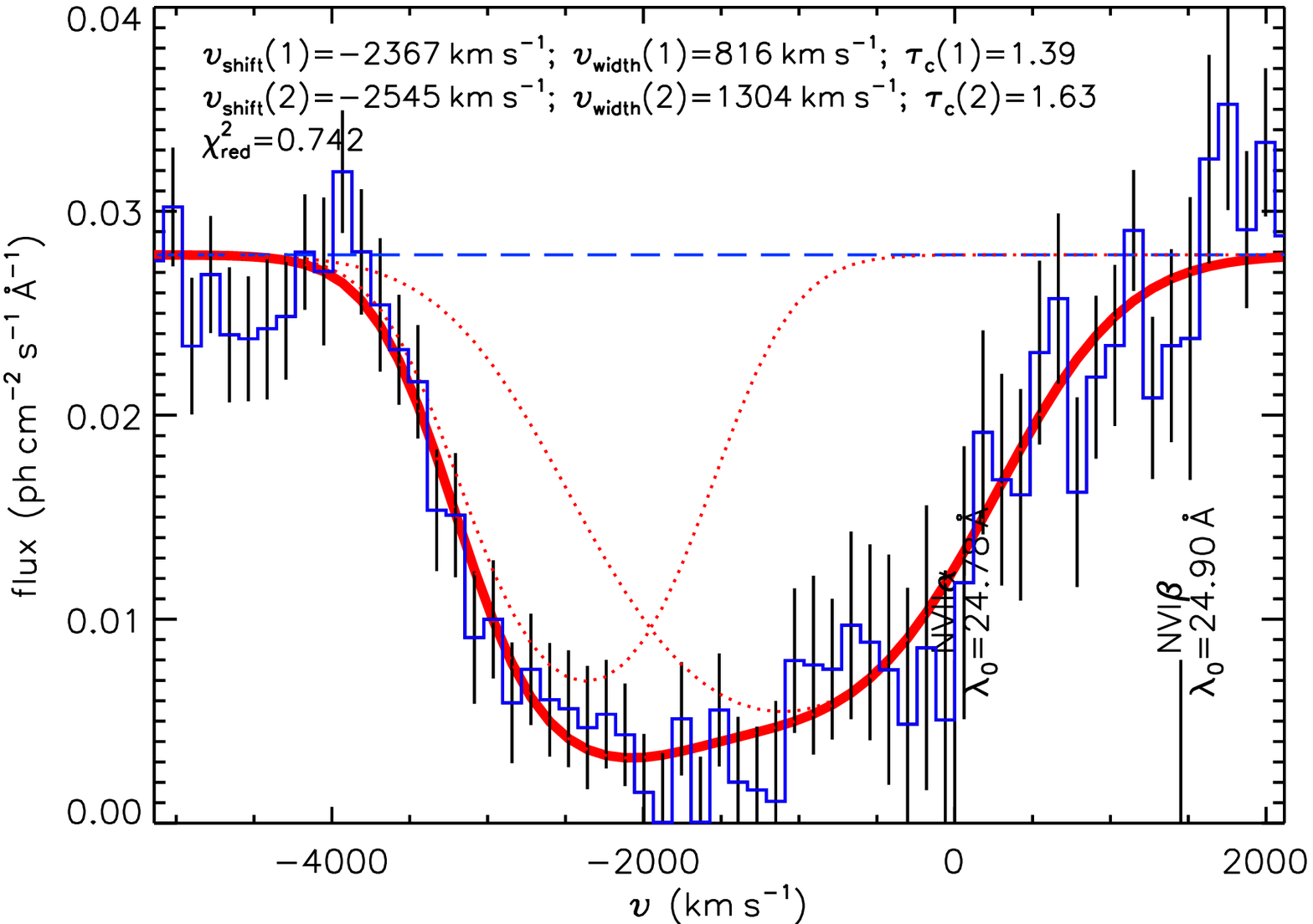}}%n76_add}\{n76_diff}
\caption{\label{fig:n7}Absorption line profile fitting to the bright spectrum
(blue histogram) following the concepts described by \cite{nessv2491} for the
two blended lines of N\,{\sc vii} ($\lambda_0=24.78$\,\AA) and
N\,{\sc vi} ($\lambda_0=24.90$\,\AA); labels are included in the
rest-frame of the N\,{\sc vii} line with black vertical lines.
The respective best-fit models are shown with a thick red line,
the assumed continuum level with the dashed blue line, and the model
components for each line with dotted red lines.
{\bf Top left}: For comparison, \chandra\ spectra of RS\,Oph taken on
days 39 (ObsID 7296, light grey) and 67 (ObsID 7297, dark grey) are
included, rescaled to the assumed continuum level showing the absorption
line in V3890\,Sgr was much broader.
{\bf Top right}: Only a single line is fitted assuming that all contributions
come from the N\,{\sc vii} line.
{\bf Bottom left}: Assuming, the emission during the dip was also present
during the bright phases, the dip spectrum is added to the model in each
iteration step. The thin red dotted line represents the model spectrum without
the addition of the dip spectrum.
{\bf Bottom right}: Model fit to the difference spectrum in the
right panel of Fig.~\ref{fig:diff}.
The fit results of all lines are
given in Table~\ref{tab:lprof}, see \S\ref{sect:analysis:lprofiles} for discussion.
}
\end{figure*}

In Fig.~\ref{fig:n7}, four options are shown illustrating how to
deal with the blended
N\,{\sc vii}\,Ly\,$\alpha$ (1s-2p) and N\,{\sc vi}\,He\,$\beta$ (1s-3p)
lines. In the top left panel, a free fit of two line templates to the bright
spectrum is shown yielding good reproduction of the general shape of the
broad absorption feature. The N\,{\sc vi} absorption line comes out
about half as broad as the N\,{\sc vii} line. If the dip spectrum
(dark green shaded) represents diffuse emission that is always present,
the strong emission lines at
the respective rest wavelengths of N\,{\sc vii} and N\,{\sc vi} could
fill the red wing of the absorption feature which
would then mostly affect the blue-shifted N\,{\sc vi} line.
In the bottom left panel of Fig.~\ref{fig:n7}, the same two lines are
fitted to the difference spectrum between bright and dip spectra
(see right panel of Fig.~\ref{fig:diff}), and the N\,{\sc vi} line
is now much broader while the N\,{\sc vii} line is narrower. A similar
result is found when adding the model to the dip spectrum while fitting
to the bright spectrum (bottom right panel). Note that the formal value
of reduced $\chi^2$ is much smaller which is only partially owing to better
fit. To properly account for the combined uncertainty of two data sets
(bright and dip spectra), the error bars in each spectral bin are
increased, and this may be an over-estimate.
Finally, we also tested the possibility of a single line with the results
shown in the top right panel yielding a visually good fit, although
$\chi^2$ is slightly worse. The line blend is thus not well resolved.\\

As a comparison to RS\,Oph, we include in the top left panel
two \chandra\ spectra taken at different times with grey shadings.
One can clearly see that the overall absorption feature was
much narrower in RS\,Oph. While the red wing essentially
matches, the blue wing extends to much larger velocities
in V3890\,Sgr. On day 39 of the evolution of RS\,Oph (light grey),
there was also an emission line component \citep{nessrsoph},
but for RS\,Oph we were not as lucky to separately see diffuse
emission originating from outside the SSS atmosphere.\\

\begin{figure*}[!ht]
\resizebox{\hsize}{!}{\includegraphics{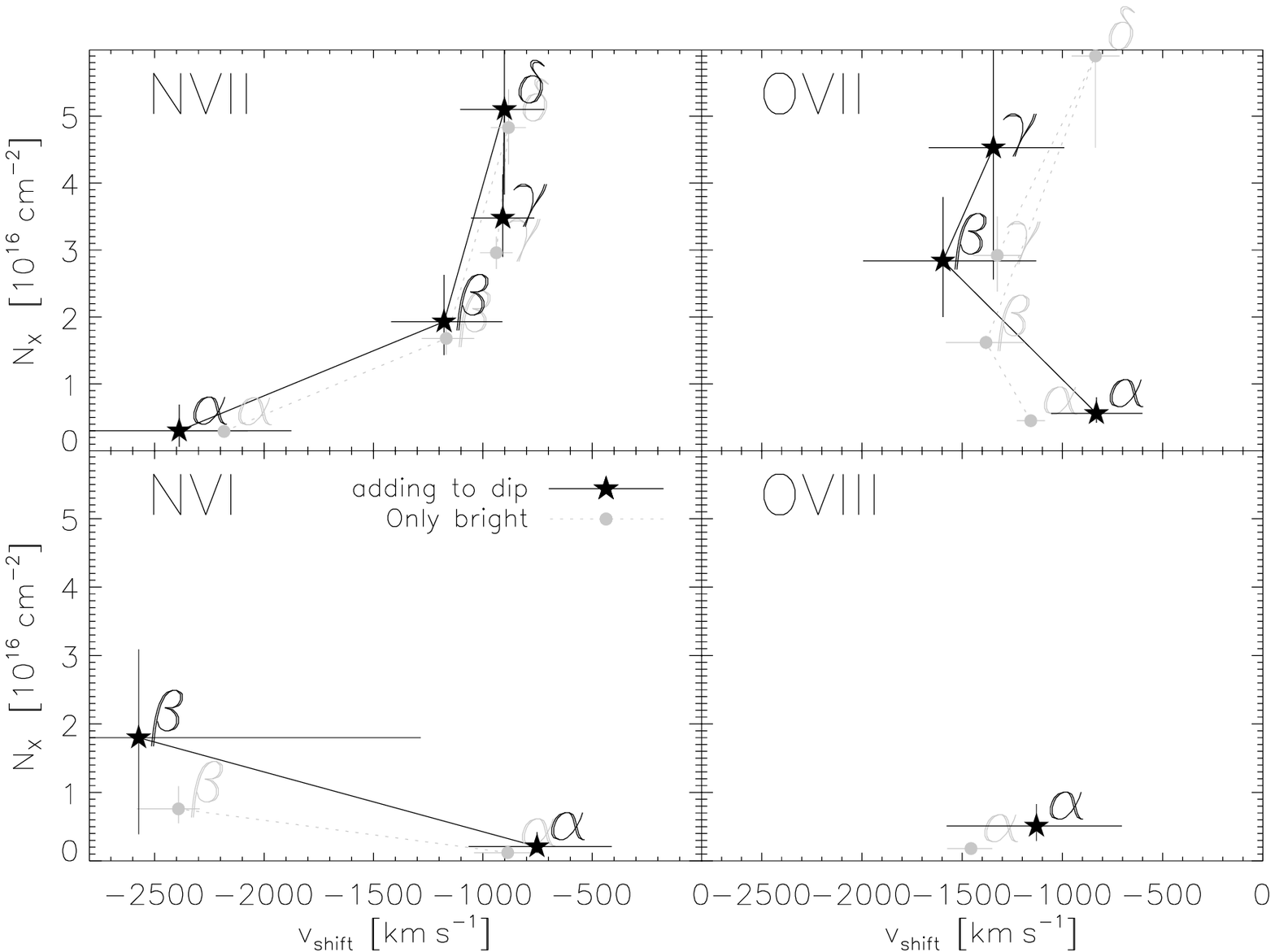}\includegraphics{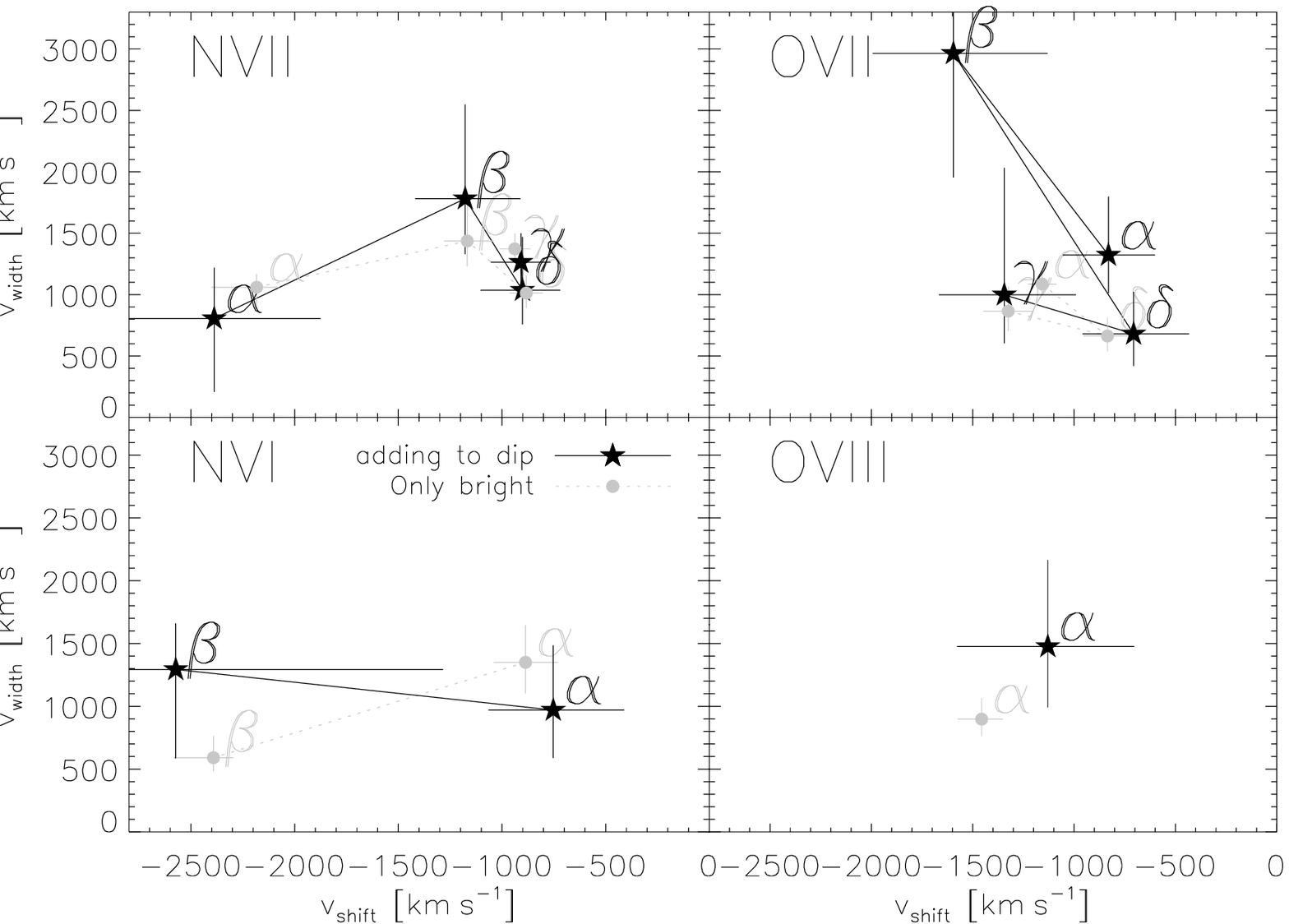}}%lshifts_nx}\{lshifts}
\caption{\label{fig:lshifts}Results from absorption line profile fitting
showing in the left panel the line column densities, and in the right panel the
line widths, both as a function of line shift. The black data points represent
the results when assuming the dip spectrum to represent a 'background' source
of the level emission observed during the dip
(see bottom left panel of Fig.~\ref{fig:n7}). The 1-$\sigma$
error bars are large because the measurement uncertainties of both,
bright and dip spectrum, are taken into account. The grey data
points represent results from simple fitting to the bright spectrum
(see top left panel of Fig.~\ref{fig:n7}). The instrumental line broadening
(of order 300\,km\,s${-1}$) has not been taken into account. Except for
some outliers (see \S\ref{sect:analysis:lprofiles}), a fairly uniform
value of line blue shift and line width can be identified of order
$800-1500$\,km\,s$^{-1}$ which is an important finding in relation to
the interpretation of the global spectral modelling results described
in \S\ref{sect:analysis:spex}.
}
\end{figure*}

We have performed the same fits to the bright spectrum while ignoring the
dip spectrum, and adding/subtracting it to/from the model to various
absorption lines and summarise all results in Table~\ref{tab:lprof}.
A graphical illustration of the results is
shown in Fig.~\ref{fig:lshifts}. It can be seen that most blue shifts cluster
around $800-1500$\,km\,s$^{-1}$ except for the N\,{\sc vii}\,$\alpha$ and
possibly N\,{\sc vi} He\,$\beta$ line (with large uncertainty).
It is also noteworthy that the 1s-2p transitions of N\,{\sc vii},
O\,{\sc vii} and possibly N\,{\sc vi} have a lower line column density
than the transitions between higher principal quantum numbers and the ground,
labelled as $\beta$, $\gamma$, and $\delta$. Meanwhile the line widths are
mostly found within the $800-1500$\,km\,s$^{-1}$ range, thus similar widths
as shifts. The only outlier is O\,{\sc vii} He-$\beta$ which is complicated
to fit as the bright spectrum does not show any clear absorption feature.
Whether or not there is an O\,{\sc vii} He-$\beta$ absorption line depends
on the assumption that the diffuse emission line feature in the dip spectrum
balances with a photospheric absorption line.\\
 
From the individual line profile analysis we can conclude that there is
a bulk velocity component of $\sim 1000$\,km\,s$^{-1}$ with a similar spread
in velocities (line widths). The application of a global model assuming a single
outflow component is thus justified (see \S\ref{sect:analysis:spex}).

\begin{table*}
\begin{flushleft}
\renewcommand{\arraystretch}{1.3}
\caption{\label{tab:lprof}Line profile fitting results with columns ion
name plus rest wavelength, line shift, line width, optical depth, oscillator
strength, and ion column density. Additional rows marked with {\it add}
indicate that the model was added to the dip spectrum (see example
in bottom right panel of Fig.~\ref{fig:n7}), and {\it diff} indicates
that the absorption model was fitted to a difference spectrum (see example
in bottom left panel of Fig.~\ref{fig:n7}).
}
\begin{tabular}{lccccc}
 & $v_{\rm shift}$ & $v_{\rm width}$ & $\tau$&$f$&$N_{\rm X}$\\
Ion ($\lambda_0$/\AA) & km\,s$^{-1}$ & km\,s$^{-1}$ & & & $10^{16}$\,cm$^{-2}$\\
%\hline
%\input{../0821560201/rgs_max/tab.tex}
\hline
\input{tab.tex}
\end{tabular}
\renewcommand{\arraystretch}{1}
\end{flushleft}
\end{table*}

\subsection{Timing Analysis}
\label{sect:analysis:timing}

For the interpretation of the X-ray light curve (Fig.~\ref{fig:xmmlc}),
we first apply an eclipse model assuming clumps crossing the line of sight
as applied to the X-ray light curve of U\,Sco \citep{nessusco} in
\S\ref{sect:analysis:ecl}. We then investigate for the presence of any
periodic patterns in the light curve in \S\ref{sect:analysis:psd}.

\subsubsection{Eclipse Modelling}
\label{sect:analysis:ecl}

In this section we assume obscuring bodies cause the large dip and
the various smaller dips observed in the X-ray light curve (see
Fig.~\ref{fig:xmmlc}).
The spectrum during the central deep dip indeed suggests a total eclipse
as all soft photospheric emission has disappeared.
Following the approach by \cite{nessusco} for U\,Sco, we assume
for the modelling that the various dips in the light curve are caused
by eclipses by spherical clumps orbiting around the white dwarf in
circular orbits. Since obscuring bodies may be forming and disappearing,
these assumptions may appear oversimplified, however the derived parameters
of clump radius and orbital radius also correspond to the size of the
clump and the distance from the white dwarf, even if not orbiting
around the white dwarf. For the sake of simplicity, we assume an
inclination angle of 90 degrees arguing that the inclination
angle is irrelevant for our purposes as we are only studying
dips caused by clumps crossing the line of sight while there may be
many other clumps in the system.\\

If the central deep dip was caused by an orbiting body, it is tempting
to interpret the initial rise in X-rays as an eclipse egress caused by
the same body orbiting at a period of $\sim 13.5$\,ks. However, if this
period is stable, the next deep dip would then start its ingress still
within the observation which
is not observed. The initial rise to maximal SSS emission thus needs to be
attributed to an additional body while the central dip could be associated
with a large obscuring clump of orbital period~$P > 13.5$\,ks. We note,
however, that the \asat\ light curve after the \xmm\ observation
contains no dips (see Fig.~\ref{fig:mmlc}). The deep dip in the
\xmm\ observation fell exactly between two \asat\ data points, and we have
not found any significant periodicity around 13.5\,ks in the \asat\
light curve. It appears
unlikely that all dips from an orbiting large clump would escape coverage
by \asat. The occulting body having caused the large dip may have thus
disappeared, or the next orbit may have been above or below the sight line.\\

A gravitationally bound system with a large body orbiting at
$P > 13.5$\,ks would be
equivalent to $a>1.23 \cdot ({M_{WD}/M_\sun})^{1/3} \cdot R_\sun$, with
$a$ the semi-major axis of the clump orbit, and $M_{WD}$ the white
dwarf mass. We find the obscuring body in a wide orbit around the white
dwarf primary, with $a > 1.23\cdot R_\sun$ for a massive white dwarf
$M_{WD} >  M_\sun$, which is expected for a recurrent nova system with
a recurrence timescale as short as V3890\,Sgr. From the total eclipse
duration one can estimate the size for this obscuring body to be
$r \approx 2.8 \cdot R_{WD}$, with $R_{WD}$ the white dwarf photospheric
radius. This same result for the clump radius is recovered by fitting
the ingress and egress light curve shapes, from which one can also find
the tangential velocity for the obscuring body as
%$v_T = 1.8 \cdot 10^{-3} \cdot R_{WD} \ s^{-1}$.
$v_T = 1.3 \cdot 10^{6} \cdot (R_{WD}/R_\sun)\ ms^{-1}$. 
For a circular orbit, this corresponds to an orbital radius of
%$a=0.12 \cdot R_{WD} \cdot (R_{WD}/R_\sun)^{-3} \cdot (M_{WD}/M_\sun)$,
$a=0.12 \cdot (R_{WD}/R_\sun)^{-2} \cdot (M_{WD}/M_\sun) \cdot R_\sun$,
which results in $a=161 \cdot R_{WD}$ when assuming a white dwarf mass
$M_{WD}=1.35\cdot M_\sun$ close to the Chandrasekhar mass limit and
a photospheric radius $R_{WD}=0.1 \cdot R_\sun$ (see \citealt{page20}).

\begin{figure}[!ht]
 \resizebox{\hsize}{!}{\includegraphics{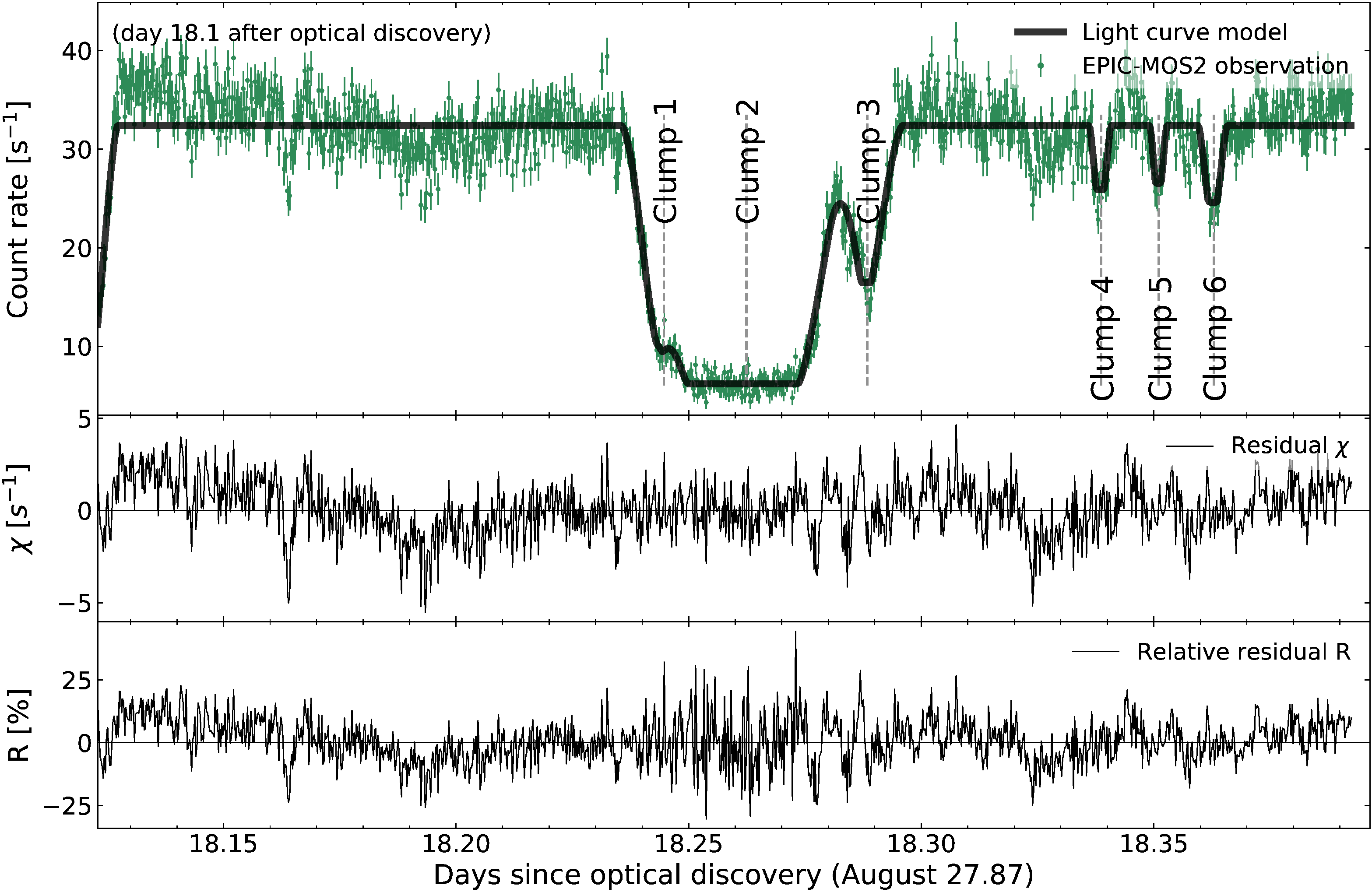}}%eclipse_model_EMOS2.ps
 \caption{Eclipse modelling for the \emph{XMM-Newton} MOS2 X-ray light curve data of V3890\,Sgr, with model parameters provided in Table~\ref{tab:eclipse_model}; see \S\ref{sect:analysis:ecl} for discussion.
 }
 \label{fig:eclipse_model}
\end{figure}

The observed X-ray (MOS2) light curve can be reproduced by a model including a
compact white dwarf photosphere at a count rate of $26.20$ counts per
second, in addition
to faint shock emission originating from a more extended region at
$6.17$\,counts per second (see \S\ref{sect:descr:dip}). The bright
photospheric emission
is dynamically obscured by multiple orbiting clumps, as the additional
substructure in the central eclipse can only be modelled by introducing
two smaller clumps next to the large orbiting body, with clump sizes as
derived from the dip-depths. The eclipse modelling results are shown in
Figure~\ref{fig:eclipse_model}, with the best-fit parameters provided in
Table~\ref{tab:eclipse_model}. The egress during the initial $420~s$
could be included in the model as an extra body (i.e. clump~0 in
Table~\ref{tab:eclipse_model}), however these clump parameters are
poorly constrained as the corresponding eclipse is not fully observed.
Three additional clumps could be introduced to reproduce the three dips
following the central eclipse. However these dips could also be
interpreted as oscillations linked to the white dwarf's spin period,
see \S\ref{sect:analysis:psd}. The best-fit model parameters provided
in Table~\ref{tab:eclipse_model} face slight degeneracies in relation
to the inclination of the orbital plane and the eccentricity of the
orbit for each body, and so they do not form a unique solution.
However, this model serves as a proof of concept for the
straightforward interpretation of orbiting clumps causing the observed
structure in the soft X-ray light curve.\\

The origin of the clumps could be similar as in U\,Sco, i.e., we may
be dealing with an accretion stream as part of a reforming accretion
disk. The material within the stream may be clumpy. If the UV emission
originates from the inner regions as in U\,Sco (evidenced by reduced
UV emission during the primary eclipse), the absence of any corresponding
clump occultations in the UV (\S\ref{sect:descr:lc}) puts a strong
constraint on the fabric of the clumps being of gas nature, transparent
to UV while opaque to X-rays. Alternatively, the UV emission may originate
from outside the inner regions or even be emitted by the clumps themselves,
illuminated by the soft X-rays.\\

Since in this model, each clump is only seen once, the scenario of
clump occultations is plausible but hard to confirm.

\begin{table}

\caption{Best--fit parameters corresponding to the eclipse modelling in Figure~\ref{fig:eclipse_model}. Clump~0 models the initial egress but is poorly constrained. Note how the orbital radius $a$ for each clump depends strongly on the assumed photospheric radius $R_{WD}$.}             % title of Table
\label{tab:eclipse_model}      % is used to refer this table in the text
\centering                          % used for centering table
\begin{tabular}{c c c c}        % centered columns (4 columns)
\hline\hline                 % inserts double horizontal lines
No. & Radius $r$ & Orbit $a$ & Phase $\phi$ \\    % table heading 

 & $\bigg[\frac{R_{WD}}{0.1 \cdot R_\sun} \bigg]$ & $\bigg[\Big(\frac{R_{WD}}{0.1 \cdot R_\sun}\Big)^{-2} \frac{M_{WD}}{1.35 \cdot M_\sun}\bigg]$& (day) \\

\hline                        % inserts single horizontal line
   $0$ & $0.88 \cdot R_\sun$ & $\, \, \, 3.4 \cdot R_\sun$ & $18.099$ \\
   $1$ & $0.09 \cdot R_\sun$ & $\, \, \, 7.5 \cdot R_\sun$ & $18.244$ \\
   $2$ & $0.29 \cdot R_\sun$ & $15.2 \cdot R_\sun$ & $18.262$ \\
   $3$ & $0.08 \cdot R_\sun$ & $\, \, \, 7.5 \cdot R_\sun$ & $18.288$ \\
   $4$ & $0.05 \cdot R_\sun$ & $\, \, \, 1.0 \cdot R_\sun$ & $18.339$ \\
   $5$ & $0.05 \cdot R_\sun$ & $\, \, \,0.6 \cdot R_\sun$ & $18.351$ \\
   $6$ & $0.05 \cdot R_\sun$ & $\, \, \, 1.6 \cdot R_\sun$ & $18.362$ \\
\hline                                   %inserts single line
\end{tabular}
\end{table}

\subsubsection{Power Spectrum Analysis}
\label{sect:analysis:psd}

In the top two panels of Fig.~\ref{fig:xmmlc}, 22 vertical tick marks are included
that are separated by 18.1 minutes. Each of the smaller dips in all X-ray light
curves, plus the in- and egress parts of the dips occurs at one of these marks suggesting
a quasi periodic behaviour. On the other hand, the majority of tick marks (14/22=64\%) do not
match any particular feature in the light curve.
To substantiate this visual impression, in this section we perform formal timing
analysis using the Lomb-Scargle method \citep{scargle1982}. The OM UVM2 light
curve contains no periodic signal, and we thus focus on the X-ray timing
analysis. The shaded areas in the top panel of Fig.~\ref{fig:xmmlc} indicate
three time segments that we analyse separately, the bright episode before the
dip, the dip itself, and the bright episode after the dip. One can see that
the variability behaviour changed after the dip showing more smaller dips than
before the deep dip.\\

\begin{figure}
\resizebox{\hsize}{!}{\includegraphics{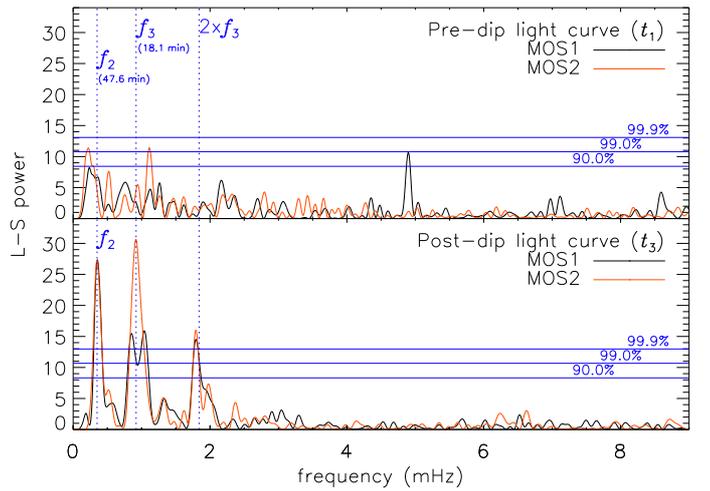}}%psd
\caption{Periodograms of the detrended light curve segments $t_1$ (pre-dip, top panel)
and $t_3$ (post-dip, bottom panel) computed from the respective MOS1 (black)
and MOS2 (orange) data. Vertical blue dotted lines represent the detected
frequencies $f_2=0.35$\,mHz and $f_3 = 0.92$\,mHz with its first harmonics
$2\times f_3\sim1.84$\,mHz (see Table~\ref{tab:periods}).
Horizontal blue lines represent the 90, 99 and 99.9\% confidence levels.}
\label{pds_3}
\end{figure}

Each time interval was
de-trended with a 3rd order polynomial before computing the corresponding
periodograms. For the time interval $t_2$, during the dip, no periodic
signal was found, and we thus focus on the pre- and post-dip time intervals,
$t_1$ and $t_3$, respectively. In Fig.~\ref{pds_3}
we show the resulting periodograms. The pre-dip periodogram contains
some peaks rising above the 90\% confidence level, however, none of them is
detected in both MOS1 and MOS2 light curves; see Table~\ref{tab:periods}.
We thus conclude that there is no significant periodic signal in the pre-dip
light curve.\\

\newcommand*{\movedown}[1]{%
  \smash{\raisebox{-1.5ex}{#1}}}
\begin{table}
\caption{Frequencies derived from periodogram peaks with amplitude above
99\% confidence level. Errors are estimated as half width at half maximum
of the power curve. Comparable values are aligned. Since the value of
0.92\,mHz measured in the t$_3$ interval of MOS2 agrees within the errors
with both $f_3$ and $f_4$, it is placed in the middle.}
\begin{center}
\begin{tabular}{lcccc}
\hline
\hline
& $t_1$ MOS1 & $t_1$ MOS2 & $t_3$ MOS1 & $t_3$ MOS2\\
\hline
$f_1$ & -- &$0.23 \pm 0.07$ & -- & --\\
$f_2$ & -- & -- & $0.36 \pm 0.05$ & $0.35 \pm 0.05$\\
$f_3$ & -- & -- & $0.85 \pm 0.06$ & \movedown{$0.92 \pm 0.09$}\\
$f_4$ & -- & $1.11 \pm 0.05$ & $1.04 \pm 0.07$ &\\
$f_5$ & -- & -- & $1.80 \pm 0.10$ & $1.80 \pm 0.05$\\
$f_6$ & $4.90 \pm 0.05$ & --& -- & --\\
\hline
\end{tabular}
\end{center}
\label{tab:periods}
\end{table}

Meanwhile, the post-dip light curve contains three strong peaks in the
power spectrum at frequencies below 2\,mHz. 
The lowest-frequency period of 47.6 minutes ($f_2=0.35$\,mHz) is detected
consistently in both MOS1 and MOS2 light curves. This corresponds to a
period of 2857\,s but the post-dip light curve only contains
3 cycles of variations and we do not consider this sufficient evidence
thus discarding this period from further studies. A period of 18.1 minutes
($f_3=0.92$\,mHz) appears as a single peak in the MOS2 light curve while
split into two signals at $f_3=0.85$\,mHz and $f_4=1.1$\,mHz and lower
L-S power in the MOS1 light curve. The lower L-S power indicates a smaller
amplitude, and that can be due to pile up. MOS1 was operated in Small
% A possible cause for observing a single frequency as two
%peaks in a power spectrum may be amplitude variations as discussed by
%\citet{dobrness17}. The MOS1 was operated in Small
Window mode that has a longer readout time than the Timing mode that was
used for the MOS2. The longer readout time for MOS1 can lead to stronger
pile up effects: two photons arriving within the readout time are registered
as one photon with the sum of their energies. The distortion of the
spectrum is the most critical effect but it can also have an impact on the
light curve as times of higher count rates suffer more pile up and slight
reduction of count rate compared to times of lower count rates. That can
reduce peaks in the light curve leading to a smaller amplitude in MOS1
compared to MOS2.
%, and a closer look at the top two panels of Fig.~\ref{fig:xmmlc} indeed reveals that the MOS2 light curve appears more structured with sharper peaks.
 We thus suspect the less significant peak in the MOS1 periodogram is a
consequence of pile up, and both light curves are consistent with a frequency
of 0.92\,mHz, thus the 18.1-minute cycle marked in Fig.~\ref{fig:xmmlc}.
The frequency $f_5$ is about twice of 0.92\,mHz, and we
interpret this peak as the first harmonics of the 0.92\,mHz frequency.
Therefore, we conclude that the value of the 0.92\,mHz is the fundamental
frequency, thus confirming the visual impression described above.
While it is only detected in the light curve segment after the
major dip, there is one smaller dip at day $\sim 18.153$ which coincides
with one of the 18.1-minute tick marks and may thus belong to the same
cyclic behaviour.\\

\begin{figure}
\resizebox{\hsize}{!}{\includegraphics{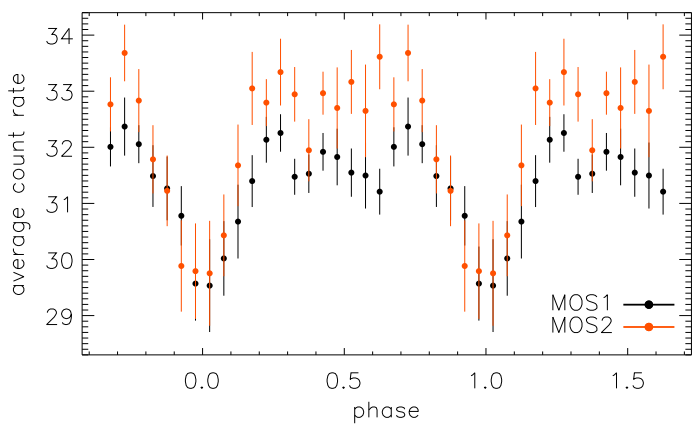}}%phase
\caption{Phased MOS1 and MOS2 light curves for the post-dip episode,
$t_3$, assuming the frequency of 0.92\,mHz. The points represent averaged
values with error of the mean.}
\label{fig:phase}
\end{figure}

In Fig.~\ref{fig:phase} we show phased and binned MOS1 and MOS2 light curves.
The amplitude of the variations is of order 12\% with a slightly higher
amplitude in the MOS2 light curve. As discussed above, this is likely due to
pile up in the MOS1 detector, owing to the different mode.\\

As independent evidence for the 18.1-minute period, we searched for this
period in the \asat/SXT data but have not detected it. This weakens the evidence,
however, we caution that a 10\% amplitude of brightness variations is close
to the limit of what can be detected with \asat.

\subsection{Spectral modelling}
\label{sect:analysis:spex}

While the analysis in the previous section is closest to the data, we now
attempt a global approach in which all atomic data are used in a
self-consistent way based on the respective physical assumptions given in
the following sub sections.
The phenomenological spectral modelling was performed using the SPEX\footnote{\href{https://var.sron.nl/SPEX-doc/}{https://var.sron.nl/SPEX-doc/}} code
\citep{spex,spex2} following the approach by \cite{pinto12} for V2491\,Cyg.

\subsubsection{Spectral modelling of the dip spectrum}
\label{sect:analysis:spexdip}

The observational evidence suggests the dip spectrum to be representative
of a plasma located far enough outside of the atmospheric SSS emission to
escape any occultations such that it is always present. The best candidate
for this emission component are shocks whose emission has been seen
on day 6.3 with \chandra\ to be consistent with a collisional plasma and
which we have shown to have cooled (Fig.~\ref{fig:shock}).
Emission lines arising from ions formed at lower temperatures (such as
O\,{\sc vii}) can thus be expected to be stronger than in the \chandra\
observation on day 6.3 \citep{orio2020}. These 'cooler' lines arise
longwards of 18\,\AA\ where the bright SSS emission is observed.\\

\begin{figure}[!ht]
\resizebox{\hsize}{!}{\includegraphics{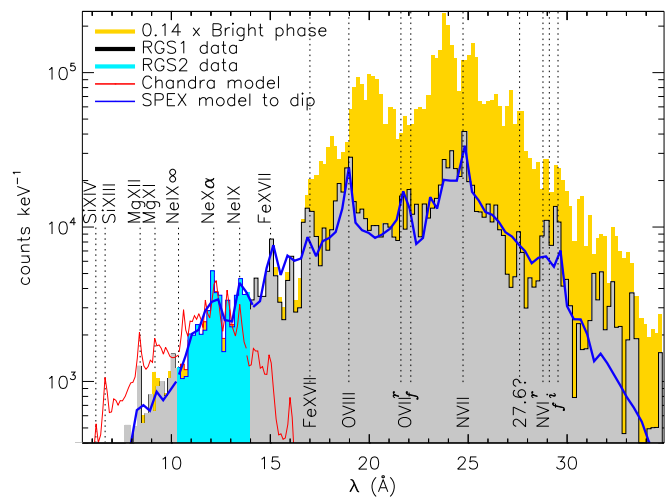}}%v3890sgr_dip.eps
\caption{\label{fig:dipspecfit}Comparisons of spectral models with the dip
(grey) and bright (yellow) spectra. The red line is the best-fit model to the
earlier \chandra\ spectrum on day 6.3 \citep{orio2020}, folded through the
RGS response, and one can see that it has more hard emission and less soft
emission than the dip spectrum on day 18.12.
The blue line indicates the SPEX model for the dip spectrum, see \S\ref{sect:analysis:spexdip} for discussion.
}
\end{figure}

In Fig.~\ref{fig:dipspecfit} we show dip and bright spectra in log units to
emphasise the fainter end of the emission. The red line is the best-fit
APEC model to the earlier \chandra\ spectrum \citep{orio2020}, folded through
the RGS response without any rescaling. It reproduces well the range
around the Ne\,{\sc x} line of the (grey) dip spectrum on day 18.12 but is
brighter at shorter wavelengths and fainter at longer wavelengths. This
indicates the cooling of the shocked ejecta. The yellow
shaded area is the (down-scaled) bright spectrum, and it can clearly be
seen that it deviates longwards of 17\,\AA. The blue curve is a best-fit model
obtained with SPEX which reproduces well the faint as well as
the brighter parts of the dip spectrum.\\

With SPEX, we have tested models including multiphase plasma in either
collisional or photoionisation equilibrium.
SPEX provides both a large database of atomic data and a powerful data
fitting package based on the {\scriptsize{CSTAT}} parameter as a goodness of fit criterion.
The {\scriptsize{CSTAT}} parameter scales well with $\chi^2$ and it does not depend on the assumption
of Gaussian noise in the data. It assumes Poissonian statistics and is thus much more accurate if
there are bins $<25$ counts.
For a reference to the accuracy and power of {\scriptsize{CSTAT}} and its application to SPEX,
see \cite{kaastra2017}.\\

We describe the X-ray continuum with a blackbody component, {\scriptsize{BB}}. The
absorption from the cold interstellar medium located along the line of sight towards
V3890\,Sgr and/or in its circumstellar medium is described with the {\scriptsize{HOT}}
model in the SPEX nomenclature with a low temperature
$0.2$\,eV (e.g. \citealt{Pinto2013}) characterised by mainly neutral species.

Line emission (and/or absorption) from a plasma in photoionisation equilibrium is
provided by the newly implemented and self-consistent
{\scriptsize{PION}}\footnote{\href{https://var.sron.nl/SPEX-doc/manualv3.05/manualse72.html}{https://var.sron.nl/SPEX-doc/manualv3.05/manualse72.html}}
 model in
SPEX, which is optimised to perform instantaneous calculation of
ionisation balance, once a photoionising continuum is provided (in this case the
blackbody), and of the corresponding transmission and emission of the photoionised
plasma. For the dip spectrum, which is dominated by emission lines, we chose the
option to include only emission lines and not transmission in order to understand
whether these can be described by gas in photoionisation balance. We found that
a single
{\scriptsize{PION}} component is not able to fit the strongest emission lines
even if assuming exotic abundances and parameters.

We have also tested models of collisionally-ionised gas with a Gaussian or a
power law-like emission measure distribution ({\scriptsize{GDEM}} and
{\scriptsize{PDEM}} models in SPEX) but again without success.
Non equilibrium models ({\scriptsize{NEI}}) also do not work, which means that
separate model components are required. Particularly difficult is to simultaneously
fit the resonance and forbidden lines of the O\,{\sc vii} triplet and the Fe\,{\sc xvii}
L-shell lines at 15-17\,{\AA}. Moreover, the observed resonant lines seem fainter
than their paired forbidden lines when compared to the model predictions. Since the
models assume an optically thin plasma, the reduced resonance line emission relative
to forbidden lines suggests resonant line scattering out of the sight line.

\begin{table*}
\caption{RGS best-fit parameters to the two flux-resolved spectra.}
\begin{center}
\renewcommand{\arraystretch}{1.1}
\small\addtolength{\tabcolsep}{+5pt}
\begin{tabular}{c | c | c | c c c}
\hline
 Component            & Parameter                    &  Low-flux dip spectrum  &  High-flux non-dip spectrum   \\
\hline
\multirow{4}{*}{Blackbody}
    & Area ($10^{18}$ cm$^2$)                        & $4.5\pm0.4$               & $4.9\pm0.2$    \\
    & Radius (km)                                    & $10800\pm1700$            & $11200\pm1360$    \\
    & k$T$ (eV/$10^5$\,K) \index{\footnote{}}        & $62\pm1$ / $7.2\pm0.1$    & $90\pm1$ / 10.4$\pm0.1$ \\
    & ${\rm L}_{\rm X}$ ($10^{36}$ erg\,s$^{-1}$)    & $19\pm 6$                 & $178\pm 16$ \\
    & ${\rm L}_{\rm bol}$ ($10^{36}$ erg\,s$^{-1}$)  & $74$                      & $320$         \\
\hline
\multirow{6}{*}{CIE\,1}
    & $n_{\rm e}$ $n_{\rm H} V (10^{58}$ cm$^{-3})$  &  $1.12\pm0.05$            & \multirow{6}{*}{All fixed}  \\
    & k$T$ (eV/$10^5$\,K)                            &  $1320\pm50$ / $153\pm6$  & \\
    & ${\rm L}_{\rm X}$ ($10^{36}$ erg\,s$^{-1}$)    &  $0.496\pm0.023$          & \\
    & ${\rm L}_{\rm bol}$ ($10^{36}$ erg\,s$^{-1}$)  &  $0.68$                   & \\
    & $\sigma_V$ (km s$^{-1}$)                       &  $890\pm90$               & \\
    & N / H                                          &  $11\pm 3$                & \\
\hline
\multirow{4}{*}{CIE\,2} & $n_{\rm e}$ $n_{\rm H} V (10^{58}$\,cm$^{-3})$         & $3.7\pm0.9$    &  \multirow{6}{*}{All fixed}  \\
    & k$T$ (eV/$10^5$\,K)                           &  $180\pm50$ / $20.9\pm6$   & \\
    & ${\rm L}_{\rm X}$ ($10^{36}$ erg\,s$^{-1}$)   &  $2.33\pm0.58$             & \\
    & ${\rm L}_{\rm bol}$ ($10^{36}$ erg\,s$^{-1}$) &  $6.2$                     & \\
    & $\sigma_V$ (km s$^{-1}$)                      &  $890$ (coupled)           & \\
    & N / H                                         &  $11$  (coupled)           & \\
\hline
\multirow{6}{*}{PION}& $N_{\rm H}$ (10$^{21}$ cm$^{-2}$)   & $\lesssim0.6$   & $ 18\pm3$      \\
    &  Log $\xi$ (erg\,s$^{-1}$ cm)  & $ \equiv 2.29 $                     & $ 2.29\pm0.05$ \\
    & $\sigma_V$     (km s$^{-1}$)   & $ \equiv 350  $                     & $ 350\pm5 $       \\
    & $\Delta_V$     (km s$^{-1}$)   & $ \equiv -900 $                     & $-900\pm10$     \\
    &  N / H                         & $ \equiv 60 $                       & $ 60\pm 20 $       \\
    &  O / N                         & $ \equiv 0.7 $                      & $ 0.7\pm0.1 $      \\
    &  C / N                         & $ \equiv 0.11 $                     & $ 0.11\pm0.01 $   \\
\hline
\multirow{2}{*}{Cold gas}   & $N_{\rm H}$ (10$^{21}$ cm$^{-2}$) &  $4.2\pm0.1$    &  $4.3\pm0.1$      \\
    &   O / H                               & $ \equiv 1 $      &  $1.1\pm0.1$      \\
\hline
\multirow{1}{*}{Dust}   &O {\scriptsize{I}} (10$^{17}$ cm$^{-2}$)  &  $\lesssim5$  &  $8.1\pm0.4$   \\
\hline
Statistics    &    $\chi^2$ / dof           &  2603/1685=1.5     &  10070/1716=5.9   \\
 \hline
\end{tabular}
\label{table:rgs_fit_colors}
\end{center}
Notes. Uncertainties are given at 1-$\sigma$ (68.3\%) level.
The {\scriptsize{CIE}} parameters are only fitted for the emission-line dominated low-flux dip spectrum. The X-ray luminosities (${\rm L}_{\rm X}$) are computed between
0.3--10 keV (assuming a distance of 9\,kpc, $L_{\rm bol}$
from 0.01--100 keV ($L_{\rm bol}$ are extrapolated and, therefore, provided without error bars). $\sigma_V$ and
$\Delta_V$ refer to the velocity dispersion and the line-of-sight velocity (Doppler shift), respectively. {\scriptsize{PION}}
photoionised absorption and dust (AMOL) absorption model parameters are also reported for the low-flux spectrum for completeness, but the low continuum prevents us from obtaining a significant detection.
\end{table*}

We achieve the qualitatively good fit illustrated with the blue line in Fig.~\ref{fig:dipspecfit}
by using an emission model consisting of two
isothermal collisionally-ionised components ({\scriptsize{CIE}} models in
SPEX nomenclature) and the parameters listed in Table~\ref{table:rgs_fit_colors}.
We have also tested a model with photo-ionised components ({\scriptsize{PIE}}) which could be
associated with the absorbing plasma in the ({\scriptsize{PION}}) component. However,
{\scriptsize{CIE}} and {\scriptsize{PIE}} fits are statistically indistinguishable.
In order to distinguish between {\scriptsize{CIE}} and {\scriptsize{PIE}},
we would need well-resolved He-like triplet lines, however, as discussed in
\S\ref{sect:descr:dip} and seen in Fig.~\ref{fig:he}, the critical intercombination
and forbidden lines are not well-enough resolved, owing to weak lines in contrast
to the continuum and the resonance line self-absorbed by an unknown amount,
further complicated by the broadening of the lines.
The weak conclusion in \S\ref{sect:descr:dip} is that during the
dip, we only see the emission originating from the shocked plasma, thus our choice
of the {\scriptsize{CIE}} components while during the bright phase we might see
additional photoexcited plasma.\\

The velocity dispersion of the two {\scriptsize{CIE}} components is tied in the
fit and yields $v=890\pm90$\,km\,s$^{-1}$ with a line-of-sight velocity consistent with zero
(i.e. at rest). The two {\scriptsize{CIE}} added contribute a total of
$\sim 10^{37}$ erg\,s$^{-1}$ within 0.3-10\,keV, assuming a distance of 9\,kpc
while the blackbody
has an X-ray luminosity of about $2\times10^{37}$ erg\,s$^{-1}$.
The abundances are broadly consistent with Solar except for nitrogen, which is a factor
$\sim10$ higher as shown by its prominent emission lines despite its cosmic abundance
being lower than oxygen. The abundance scale used here is the default one in
SPEX (proto-Solar, \citealt{Lodders2009}). Enhanced N abundance is expected
in CNO burning plasmas.\\

We caution that the error estimates in Table~\ref{table:rgs_fit_colors} are
strictly statistical errors, not taking into account systematic effects.
To get an idea of systematic effects, we have performed several tests
probing different scenarios. For example, with and without a {\scriptsize{PION}}
component and variable or fixed values of $N_{\rm H}$ and CF, we find
areas between $8.2\times 10^{15}$\,cm\,$^2$ and  $4.5\times 10^{18}$\,cm\,$^2$
and blackbody temperatures between k$T=62-125$\,eV. The parameters of the dip
spectrum are thus not as well constrained as suggested from the formal errors.
Statistically, the all-frozen configuration is not favoured which is why we report
the results with $N_{\rm H}$ as free parameter. However, there is only a difference
in {\scriptsize{CSTAT}} parameter
of 30 between a model without a {\scriptsize{PION}} absorber and {\scriptsize{PION}} 
with free $N_{\rm H}$. This means {\scriptsize{PION}} is statistically not required for the
low-flux spectrum. Therefore, it is still possible (although not necessarily the truth)
that something opaque stood in our line of sight obscuring the inner region (which was
absorbed by {\scriptsize{PION}}) but not any other (cooler) scattered emission coming from
outside and not absorbed by {\scriptsize{PION}}.

\subsubsection{Spectral modelling of the bright spectrum}
\label{sect:analysis:spexbright}

The RGS spectrum of V3890\,Sgr extracted during the high-flux intervals as indicated
in Fig.~\ref{fig:xmmlc} is characterised
by strong blue-shifted absorption lines carved into a supersoft X-ray continuum with
additional emission lines which appear primarily in the harder energy band, below
17\,{\AA}. We also attempted to model the RGS spectrum extracted during this phase with
SPEX following the approach of \cite{pinto12} for V2491\,Cyg.

As in \S\ref{sect:descr:smap}, we adopt a blackbody continuum component
({\scriptsize{BB}}) and interstellar neutral absorption ({\scriptsize{HOT}}) with
a low temperature parameter. Given
the higher continuum in this spectral state and the strength of the oxygen K edge,
we account for any dust or molecular gas absorption through the {\scriptsize{AMOL}}
model in SPEX, which was necessary to correctly model the neutral O K edge
in nova V2491\,Cyg \citep{pinto12} and in common X-ray binaries \citep{Pinto2013}.
The line emission presumably produced by shocks around the nova and likely of
collisional-ionisation nature is not included through a collisional-ionisation
model but rather by using the low-flux dip spectrum as a template emission model
that we include as an additive emission component, {\scriptsize{$F_\lambda$\,dip}}.
This choice was dictated by the fact that the emission lines do not seem to
significantly change throughout the whole observation. The photospheric absorption
lines produced in the nova ejecta are modelled with the {\scriptsize{PION}} model.
Compared to the work previously done in nova V2491\,Cyg \citep{pinto12}, this model
represents an evolution of the {\scriptsize{XABS}} model used in that paper as
{\scriptsize{PION}} calculates the ionisation balance instantaneously from the
simultaneously fitted continuum. Previously we would need to provide a given
SED and pre-calculate the ionisation balance and the strength of the lines,
thereby preventing us from obtaining constraints on both the shape of the
ionising continuum and the properties of the ejecta.\\

The spectrum extracted outside the dips is therefore modelled with a
multi-component model $F_\lambda$ symbolically described as
\begin{equation}
 F_\lambda =F_\lambda\,{\rm dip} + F_{\lambda,\mathrm{BB}} \cdot {\scriptstyle{\rm PION}} \cdot {\scriptstyle{\rm AMOL}} \cdot {\scriptstyle{\rm HOT}}
\end{equation}

Free parameters are the column density, $N_{\rm H}$, and the oxygen abundance of the interstellar
absorber ({\scriptsize{HOT}}). The strong O K edge around 23.0\,{\AA} and the 1s-2p line at 23.5\,{\AA}
allow us to achieve high sensitivity on the contributions from oxygen, dust, and gas phases thanks to the
relative ratio of the edge and line depths and the detailed shape of the edge. The column density of the
dusty {\scriptsize{AMOL}} component is a free parameter. We tested several compounds (ice, iron
oxides, silicates and some carbon molecules) and achieved the best fit using silicates from pyroxene,
although olivine provides comparable results. This agrees with previous results on nova V2491\,Cyg and
Galactic low-mass X-ray Binaries \citep{pinto12,Pinto2013}. The normalisation and the temperature of the
blackbody were also free to vary. For the {\scriptsize{PION}} component we fitted the column density,
the ionisation parameter ($\xi = L_{\rm ion} n_{\rm H}^{-1} \, R^2$), the outflow velocity, the velocity
dispersion, and the abundances of the atomic species responsible for the detected lines (e.g. C, N,
O, Si, S, Ar, Ca, Fe, Ni). We adopt Solar abundances for all other species (${\rm X/H}=1$) including
neon and magnesium because the spectral continuum is too low below 15\,{\AA} to be sensitive to their
absorption lines. This model is simpler than the one used for V2491\,Cyg in \cite{pinto12}
where we employed up to three layers of photoionised gas as there was a strong trend between the
line blueshift and the corresponding ionisation potential. As can be seen from Fig.~\ref{fig:lshifts},
the line blueshifts are rather homogeneous except for the N\,{\sc vii} line.\\

\begin{figure*}[!ht]
\resizebox{\hsize}{!}{\rotatebox{270}{\includegraphics{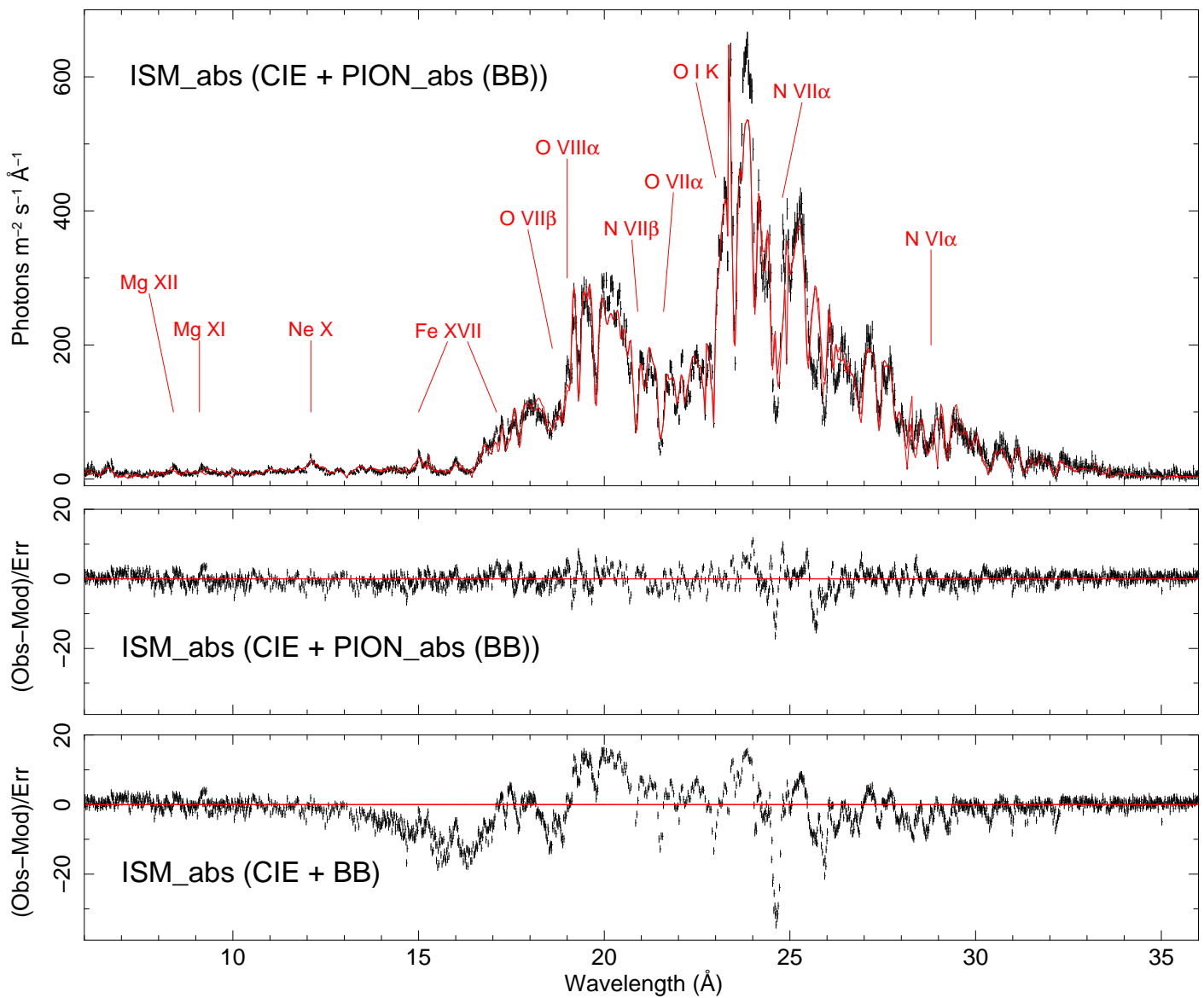}}}%spex_model
\caption{\label{fig:spex_fit}
\xmm\ RGS spectrum of V3890\,Sgr extracted
 during time intervals excluding the dip with the full SPEX model overlaid
 (top panel); for parameters, see Table~\ref{table:rgs_fit_colors}.
The rest-frame wavelengths of the dominant transitions are labelled.
 The middle panel shows the residuals, and the lower panel shows
the residuals without the {\scriptsize{PION}} model.}
\end{figure*}

The best-fit model and residuals are shown in the top two panels of Fig.\,\ref{fig:spex_fit} with
the best-fit parameters listed in the right column of Table~\ref{table:rgs_fit_colors}. Our simple
model provides a good description of the overall spectrum, matching the strengths and centroids of
most absorption lines, yielding a reduced C-Statistic of $C/d.o.f.=10070/1716=5.9$, which is a
substantial improvement over a model without {\scriptsize{PION}} absorption
($C/d.o.f.=50548/1693=30.0$) or the simple blackbody model ($C/d.o.f.=86404.84/1694=51.0$).
A reduced C-Stat
value of 5.9 is formally not considered an acceptable fit, but we notice that, given the high
count rate of the source, any uncertainties in the RGS calibration, the atomic database (whose
cross sections are known to be uncertain up to the 20\% level), the structure of the outflow
(i.e. in the case of a multiphase plasma such as in V2491\,Cyg), the presence of additional
processes like resonant scattering and photoionisation (re)emission will prevent to achieve
$C/d.o.f.\sim1$.\\

For the blackbody we estimate a temperature of $90 \pm 1$\,eV $=1\times 10^6$\,K
(similar to the SSS phase of V2491\,Cyg, \citealt{pinto12}) and an X-ray luminosity of
$178 \times 10^{36}$\,erg\,s$^{-1}$, assuming isotropy and a distance of 9\,kpc.
For the cold interstellar medium we estimate a column density
$N_{\rm H} = 4.3 \pm 0.1 \times 10^{21} {\rm cm}^{-2}$, with the oxygen abundance slightly
above the solar value and dust silicates contributing between 20--30\% of
the interstellar neutral oxygen as previously found in nova V2491\,Cyg and most X-ray Binaries
\citep{pinto12,Pinto2013}.\\

For the photoionised absorber, i.e. the nova ejecta, we estimate a column density
$N_{\rm H} = 18 \pm 3 \times 10^{21} {\rm cm}^{-2}$, an ionisation parameter
Log $\xi = 2.29 \pm 0.05\,{\rm erg\,s^{-1} \, cm}$. For the outflow velocity and the velocity
dispersion we find $-900 \pm 10$ km\,s$^{-1}$ and $350 \pm 5$ km\,s$^{-1}$, respectively.
Most element abundances
are broadly consistent with the Solar values, except for nitrogen ($N/H\gtrsim40$) and oxygen
($O/H\gtrsim25$). As previously shown for V2491\,Cyg \citep{pinto12}, the absolute abundances,
i.e. relative to hydrogen, may be subject to large uncertainties, particularly driven by the
knowledge of the exact continuum (flux and shape). However, the relative abundance ratios between
atomic species responsible for the detected lines are much better constrained
(e.g. $O/N=0.7\pm0.1$ and ${\rm C/N}=0.11 \pm 0.01$, which is expected given the lack of
strong carbon lines). The abundances obtained here have a pattern similar to the one
measured for V2491\,Cyg \citep{pinto12}.\\

We also fitted the same spectral model to the pre- and post-dip spectra separately
(time intervals in Fig.~\ref{fig:xmmlc} marked with blue and red, respectively).
We found all parameters consistent with each other except for the velocity $\Delta_V$ for
which we found $-870\pm10$\,km\,s$^{-1}$ before the dip and $-900\pm10$\,km\,s$^{-1}$
after the dip.\\

%-----------------------------Figure Start--------------------------------

\section{Summary of Results}
\label{sect:results}

In summary, we have obtained the following results directly from the observations:
\begin{itemize}
 \item The SSS emission is highly variable between two distinct brightness levels that
 we refer to as bright phase and dip phase. This short-term variability is consistent
 with \swift\ and \asat\ long-term light curves;
 \item In the UVM2 band ($\sim 1620-2620$\,\AA) no such variability is seen,
 neither on long-term nor on short-term time scales. A slow decline of 0.1\,mag per
day is seen;
 \item The X-ray spectrum consists of bright SSS emission and emission lines from the cooling
   shocked plasma;
 \item The emission lines from the shocks have evolved since a \chandra/HETGS spectrum
  was taken on day 6.3. They show weaker emission in lines formed at higher
  temperatures and equal emission from lines formed at intermediate temperatures.
  The shocked plasma has thus cooled, consistent with expectations. Consequently,
  lines formed at lower temperatures should be stronger but they arise at wavelengths
  where additionally strong SSS emission is seen. The SSS emission and the outer cooling
  shocks cannot easily be disentangled visually except possibly during the dip.
  While the \chandra/HETGS spectrum shows skewed emission line profiles, the \xmm/RGS
  resolution is not sufficient to resolve such a profile shape;
 \item The SSS continuum shape resembles a blackbody with $T_{\rm eff}\sim 6.9\times 10^5$\,K,
  and $N_{\rm H}\sim 5.4\times 10^{21}$\,cm$^{-2}$ and ionisation edges from
   O\,{\sc i} (interstellar, $N_{\rm H}\sim
   4.3\times 10^{21}$\,cm$^{-2}$), N\,{\sc vi} and N\,{\sc vii} (both local,
   $2.2\times 10^{17}$\,cm$^{-2}$ and $1.1\times 10^{17}$\,cm$^{-2}$, respectively);
 \item Blue-shifted broadened absorption lines can be identified from N\,{\sc vii}
   Ly series and O\,{\sc vii} He series. Weaker absorption features are likely also
   present from O\,{\sc viii} Ly\,$\alpha$ and He-like N\,{\sc vi};
 \item Absorption features seen at 26.93\,\AA, 30.42\,\AA, and 32\,\AA\ may arise from
   Ar\,{\sc xv}, Ca\,{\sc xi}, and S\,{\sc xiv}, although these identifications are
   uncertain owing to discrepancies with other expected lines.
   These lines have also been seen in other SSS spectra, e.g., SMC\,2016, V4743\,Sgr,
   and RS\,Oph. Especially the 32-\AA\ lines seems highly 'mobile' with the ability to
   shift by as much as 500\,km\,s$^{-1}$ within 2 weeks;
 \item The dip spectrum (2.7\,ks duration) contains emission lines from Fe\,{\sc xvii},
   O\,{\sc viii}, O\,{\sc vii}, N\,{\sc vii} and N\,{\sc vi};
 \item At low significance, the He-like triplet lines of O\,{\sc vii} suggest that the
   dip spectrum is dominated by collisional plasma at densities $<10^9$\,cm$^{-3}$ while
   during the bright phases, additional contributions from photoexcitations are seen.
\end{itemize}

\noindent
The direct results from observations are more robust than model-dependent results which
in turn can give us more quantitative results which are:
\begin{itemize}
  \item {\bf Line profile fitting} to individual absorption lines yields consistent
 line widths and blue-shifts in most lines that correspond to a bulk velocity component of
$\sim 1000$\,km\,s$^{-1}$.
    The line profiles are generally much more similar to each other than in V2491\,Cyg
    where at least three types of profiles were found \citep{nessv2491}.
    Nevertheless, some outliers are found, namely:
  \begin{itemize}
    \item The absorption line complex at 24.8\,\AA, N\,{\sc vii} Ly\,$\alpha$ and
    N\,{\sc vi} He\,$\beta$, yields a particularly high value of line shift while
    the line widths are consistent with all other lines;
    \item Line column densities of 1s-2p transitions appear lower than transitions
     between ground state and higher principal quantum numbers;
    \item The column densities of absorption lines are about a factor 2-5 lower than
    those of the absorption edges of N\,{\sc vi} and N\,{\sc vii};
  \end{itemize}
  \item {\bf Power spectrum analysis} of the light curve segment after the dip results
  in a significant detection of a
 18.1-minute period. Visual inspection of the whole light curve confirms several
 dips plus the in- and egress of the large dip and the initial rise to be in phase
 with this period. The duty cycle is low, with 36\% in the whole light curve and 50\%
 after the dip (including egress).
  \item {\bf A global phenomenological spectral model implemented in SPEX} was used to reproduce the
    observed dip and bright spectra.
  \begin{itemize}
    \item During the bright episodes before and after the deep dip, the spectrum can be
      reproduced with a 6-component model with 26 parameters. This is much less than the
      SPEX model fitted to V2492\,Cyg \citep{pinto12};
    \item Assuming spherical symmetry and a distance of 9\,kpc, the
    normalisation of the SPEX model implies a compact white dwarf of
    $\sim 6000$\,km radius which implies the photospheric radius to be close
    to the white dwarf surface;
   \item The bolometric luminosity $L_{\rm bol}$ during the bright episodes, assuming
     a distance of 9\,kpc, ranges mildly above the Eddington luminosity of a 1-M$_\odot$ object;
    \item The bright episode after the dip yields a slightly
      higher velocity derived from the blue shifts of the absorption lines
      compared to before the dip;
    \item The dip spectrum contains emission lines that can be modeled with
      two collisional equilibrium components and a faint blackbody component.
      The observed X-ray flux of this component is a factor
      $\sim 10$ fainter than during the bright phase while the derived bolometric
      luminosity during the dip is a factor $\sim 4$ lower. $T_{\rm eff}$
      is much lower while $N_{\rm H}$ and emitting area are consistent within the
      uncertainties. See, however, the discussion on systematics in \S\ref{sect:disc:spex}
  \end{itemize}

\end{itemize}

\section{Discussion}
\label{sect:disc}

\subsection{Variability}
\label{sect:disc:pulse}

Our results from \S\ref{sect:analysis:psd} suggest small dips seen
in Fig.~\ref{fig:xmmlc} to be aligned with an 18.1 minute period, albeit
at a low duty cycle and not supported by other data.
An 18.1-minute period would be typical for the rotation period of the
WD, however, the shape of the small dips cannot be explained by
anything that would be co-rotating on the surface of the visible
photosphere. Even small features would cause much broader phase
profiles than those seen in Fig.~\ref{fig:phase}. We discuss in this
section possible interpretations of the dips via clump occultation
or surface features, however, none of them necessarily periodic.\\

\cite{nessusco} describe how deep dips in the U Sco X-ray light curve
can be due to clumps passing through the line of sight that belong to
an accretion stream within the ecliptic plane. This could be a periodic
process which for U\,Sco could not be checked because less than an
orbit had been observed. For V3890\,Sgr, this scenario does not work
because the orbit is much longer, and the inclination angle is much
lower than in U\,Sco. The clump model described in
\S\ref{sect:analysis:ecl} thus assumes random clumps that are not
necessarily in an orbit around the white dwarf, and in this model,
the small dips as well as the deep dip would
coincidently occur in a semi-regular cycle.\\

The spectrum in U\,Sco during the deep dips shows clear signs of
scattered continuum emission which in V3890\,Sgr is much weaker
during the dip. The intensity of the remaining continuum emission
should be driven by the amount of surrounding scattering material
of which there must then be much less than in U\,Sco.\\

Deep dips leading to complete disappearance of SSS emission have also
been seen in V4743\,Sgr \citep{v4743} or V2491\,Cyg \citep{nessv2491}.
As these systems are also no edge-on systems like U\,Sco, it is
plausible to conclude that a large number of occulting bodies would be
distributed in all directions with no preference to the ecliptic plane.
Only a very small fraction of them leads to observable dips,
independent of the viewing direction.\\

More difficult is to understand why the minor dips all have similar
profiles (see Fig.~\ref{fig:phase}). In particular having all
the same depth requires some fine-tuning of all clumps needing to have
the same sizes and crossing geometries to cause the same types of
partial occultations.
One may consider partial absorption as it is reasonable to assume that
all clumps consist of the same material at similar physical conditions.
However, partial occultation should be energy-dependent, and we would see a
harder spectrum during the partial dips.\\

\begin{figure}[!ht]
\resizebox{\hsize}{!}{\includegraphics{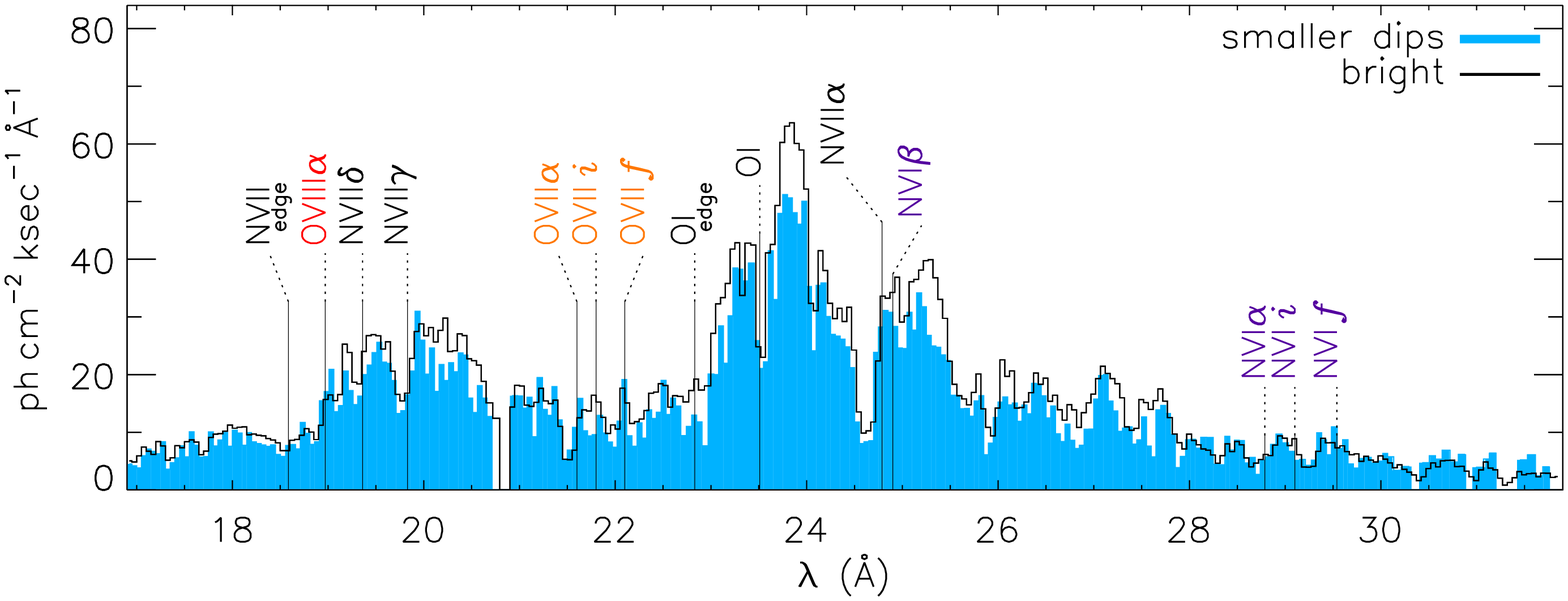}}%cmpv3890sgr_dips

\resizebox{\hsize}{!}{\includegraphics{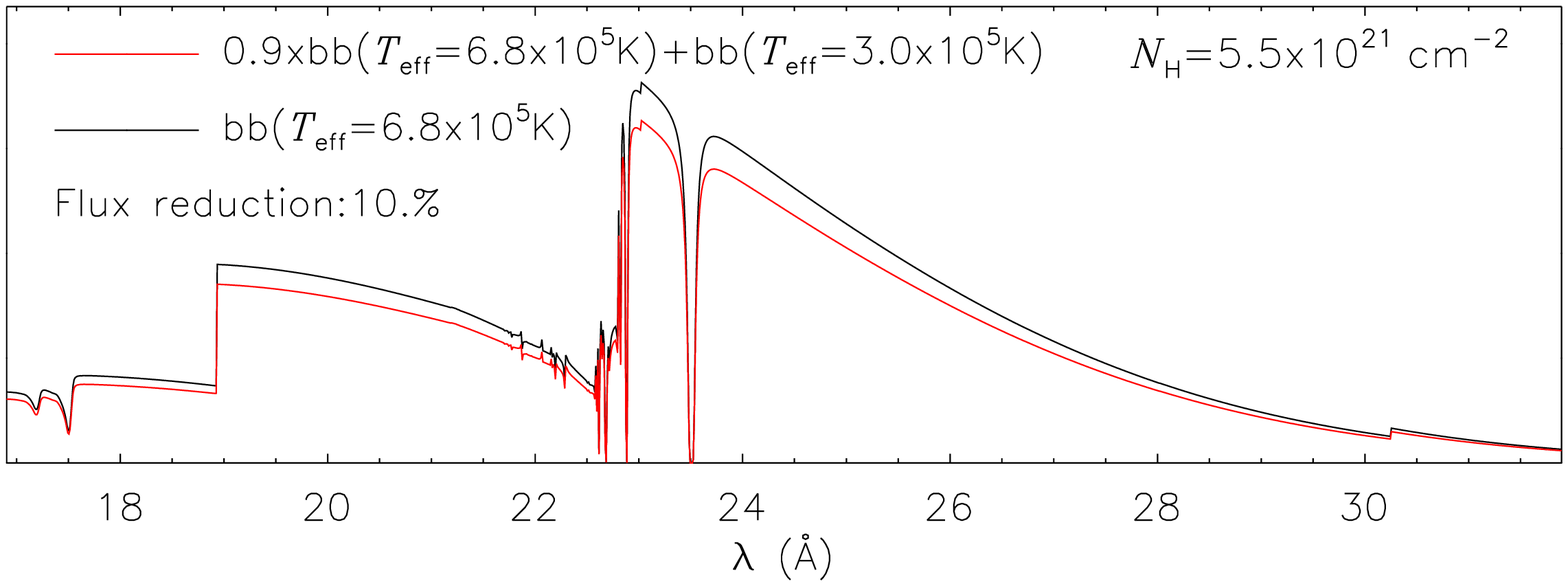}}%nhtest
\caption{\label{fig:cmpdips}{\bf Top}: Comparison of spectra
extracted during the bright intervals (black histogram)
and during three smaller dips after the deep dip
(light blue). The emission only deviates at intermediate wavelengths
but are similar at long wavelengths (driven by $N_{\rm H}$) and at
short wavelengths (driven by $T_{\rm eff}$ and edge absorption).
{\bf Bottom}: An ad-hoc model to attempt reproducing this behaviour.
The black curve represents a blackbody curve (in arbitrary
flux units) with the
best-fit parameters to the observed bright spectrum. The red
curve represents the sum of the black curve, reduced by 10\%, and
a blackbody spectrum with a lower temperature but absorbed by the
same $N_{\rm H}$ (see legend for values).
The red model could be compared to the spectrum extracted during
the smaller dips.
See \S\ref{sect:disc:pulse} for discussion.
}
\end{figure}

In the top panel of Fig.~\ref{fig:cmpdips},
the spectra during three of the minor dips is compared with the bright
spectrum, and we do see energy-dependent absorption, however, not as
expected for a partial absorber that would reduce the flux more at soft
energies (longest wavelengths) while we see the flux reduction
at intermediate wavelengths, $\sim 19-25.5$\,\AA. We have tested some simple models,
and this behaviour cannot be explained by simple absorption but requires
some more complex processes. The simplest model we could find that can
reproduce this behaviour is shown in the bottom panel of
Fig.~\ref{fig:cmpdips}. It consists of the sum of a bright spectrum
that is reduced by the same fraction (e.g. 10\%) at all
wavelengths and an additional cooler component. The sum of both components is
then absorbed by the same column density of interstellar material.\\

This model experiment suggests that the small dips may not simply be
absorption events but that a small fraction of the photosphere could
temporarily be replaced by cooler plasma.
Various possibilities could be imagined, examples may be magnetic
field lines suppressing convection like in the solar atmosphere
(sun spots),
material falling back through the photosphere creating temporarily
cooler impact craters, or small compact plasma cells temporarily
rising up from below the photosphere. If such plasma cells
are more compressed than the rest of the photosphere,
their photospheric temperature would be cooler.
In all these examples, the sum of remaining photosphere and a
smaller-size cooler component could produce dips with a spectrum
like the red model in the bottom panel of Fig.~\ref{fig:cmpdips}.\\

These are only suggestions that would require deeper studies to test
and validate which are beyond the scope of this work. Features such as
sun spots appear unlikely to us as they would co-rotate with the white
dwarf and form broader phase profiles than observed. Whether features
like sun-spots could be so short-lived that each spot creates one dip
appears rather unlikely to us. Material falling back appears possible
with each impact causing a dip, however, we cannot identify a situation
in which this would be a semi-periodic process. Meanwhile, rising plasma
cells might be controlled in some way by the rotation of the WD if some
latitudes are more favourable to such rising cells than others. However,
they would have to dissolve on time scales of the dip duration ($\sim 5$
minutes).\\

%The 18.1 minute dip repetition might be due to the WD spin period,
%with the dips due to sporadic obscuration by partial magnetically
%defined accretion curtains (as is commonly seen in intermediate
%polars), these may occur over a range of higher inclination angles
%and might be sporadic and of limited azimuthal extent from a re-forming
%accretion disc. The lack of distinctive spectral signature during the
%dips would require an exploration of photoionized absorption parameter
%space to assess the viability of this hypothesis. \\

We can therefore think of various processes causing major and minor
dips but do not have enough observational evidence that could
explain why these processes would occur periodically.

\subsection{Spectroscopy}
\label{sect:disc:spex}

Super-Soft-Source (SSS) spectra, at the time of their discovery, appeared to
be quite easy to parameterise with blackbody fits that yielded good
reproduction of observed CCD-type spectra \citep[e.g.][]{krautt96}.
The strongest reason for developing more complex models to fit CCD spectra
of SSS spectra were the unrealistically high bolometric luminosities
($L_{\rm bol}$) derived from the effective temperature
($T_{\rm eff}$) of blackbody fits. The fit quality from
fitting LTE and NLTE models has not always improved, but more
realistic bolometric luminosities ($L_{\rm bol}$) were derived
\citep[e.g.][]{balm98,parmarcal87,parmarcal83}.
Since the fit quality has not improved, the lower values of $L_{\rm bol}$
were not
required by the data but a result of expectations. The only conclusion
that can be drawn is that super-Eddington luminosities are not required
by the data, but they remain only ruled out by theory.\\

Therefore, the higher spectral resolution of the \chandra\ and \xmm\
gratings promised a breakthrough, however, the grating spectra turned out
to be notoriously complex, and no model has so far been found that yields
a statistically significant fit. A simple blackbody fit still
reproduces the general shape fairly well, although no longer at a
$\chi^2_{\rm red}\sim 1$ level. However, also no self-consistent atmosphere
model that should reproduce the resolved absorption lines has come
anywhere near $\chi^2_{\rm red}\sim 1$.\\

Visual inspection is therefore not just pedantry but is needed to conceptualise
the modelling approach. We should not only be driven by theoretical
expectations but need to take into account strategic details such as weak
forbidden lines or small line shifts that do not contribute significantly
to $\chi^2$ of a global fit. A key result from the inspection of individual
absorption lines is the finding that most lines are shifted by the same
amount of $\sim 800-1500$\,km\,s$^{-1}$. In contrast, for V2491\,Cyg,
\cite{nessv2491} found a large range of line shifts (see their figure~10),
and a global model presented by \cite{pinto12} required a large number of
layers.\\

The SPEX model presented by \cite{pinto12} may be a good compromise between
an oversimplified blackbody fit and a complex atmosphere model. Just like
an atmosphere model, it starts with a blackbody source function but then
only adds phenomenological absorbing layers rather than computing the results
of full-fledged radiative transport processes. While this approach is not
fully physically self-consistent, the full (known) atomic physics is
accounted for when fitting the
absorption lines and edges. The reproduction of the data was much better
than any atmosphere model to date, however, the number of layers was rather
large giving the impression of arbitrarily adding components
until the data are reproduced. Thanks to the prior inspection of the individual
absorption lines, this can be understood as dealing with multiple velocity
components.\\

In this work we have been able to find a model with a single absorbing layer
which is consistent with the findings from individual lines. The
{\scriptsize{CSTAT}} value of $\sim 6$ is, strictly speaking, still not an
acceptable fit but much better than anything seen before when fitting
models to SSS grating spectra. We have also
attempted fitting atmosphere models but failed to find anything close
to an acceptable fit to the data. Nevertheless, $T_{\rm eff}$ is
in rough agreement with the SPEX result.\\

From both, direct inspection and global model approaches, we can thus
conclude that, compared to V2491\,Cyg \citep{pinto12}, the atmosphere
in V3890\,Sgr appears relatively uniform. The good fit quality allows
quantitative conclusions from the resulting parameters listed in
Table~\ref{table:rgs_fit_colors}, however, always with the caveat in
mind that the conclusions are model-dependent (see below):\\

We are dealing with a hot atmosphere, close to a million degrees, much
higher than $T_{\rm eff}$ of a blackbody fit which confirms many findings
of blackbody fits underestimating $T_{\rm eff}$. This
is quite easy to understand: A model ignoring the effects of the absorbing
layers above the central source function that remove emission at the
high-energy end of the Wien tail (e.g. by ionisation edges) needs to place
the Wien tail of the source function at lower energies and
thus lower $T_{\rm eff}$. Models accounting for absorption edges will thus
always lead to high $T_{\rm eff}$. With temperature scaling with the
fourth power while radius only squared, $L_{\rm bol}$ will then be
underestimated. The inclusion of ionisation edges in a blackbody fit
is only the first step towards physically realistic model conditions.
Apparently, accounting for all (known) absorbing atomic/ionic transitions
in an arbitrary absorption layer can lead to a more adequate estimation of
$T_{\rm eff}$ of the underlying source function leading to
similar results as a full-fledged radiation transport model.
Very possibly, the exact radiation transport processes may not be the decisive
factor when finding realistic $T_{\rm eff}$ and thus $L_{\rm bol}$.\\

The resulting photospheric radius falls into the ball park of typical white
dwarf radii suggesting that the observed continuum emission originates
from close to the surface of the white dwarf. This result assumes the
emission to homogeneously be emitted by a spherically symmetric
photosphere. If the symmetry assumption is not valid, then all estimates
of radius and luminosities are invalid, and we may be dealing with much
smaller emitting regions, each of them with luminosities much lower than
$10^{38}$\,erg\,s$^{-1}$.\\

The dip spectrum, dominated by emission lines, can only be reproduced
if the presence of an additional atmosphere is assumed, modeled in SPEX with a
blackbody source function and the same {\scriptsize{PION}}
absorption layer with the same
parameters obtained from the bright spectrum. The X-ray flux of this component
has decreased by a factor $\sim 10$ while $L_{\rm bol}$ has only
decreased by a factor 4. Since the radius remained the same, the reduction
in $L_{\rm bol}$ is thus owed only to the reduction of $T_{\rm eff}$
\footnote{With values from Table~\ref{table:rgs_fit_colors},
$L_{\rm bol,dip}/L_{\rm bol, bright}=0.23=(T_{\rm dip}/T_{\rm bright})^4=0.7^4=0.23$}.
The reduction of $L_{\rm X}$ by
a factor 10 is due to the fact that the peak of the spectral energy
distribution has shifted into the FUV where a larger fraction of flux is
lost to the interstellar medium by means of photoelectric absorption by
neutral hydrogen. A quick check comparing blackbody spectra with
$T_{\rm eff}=10.4\times 10^5$\,K and $7.2\times 10^5$\,K (yielding a factor 4
in $L_{\rm bol}$) yields a factor 22 in $L_{\rm X}$ assuming
$N_{\rm H}=4.25\times 10^{21}$\,cm$^{-2}$. Therefore, at least within
the statistical uncertainties, the deep dip could
be explained by a reduced temperature alone without any other factor
such as occultations that are assumed in \S\ref{sect:analysis:ecl}.\\

The interpretation of the dip by reduced $T_{\rm eff}$ alone
and a corresponding reduction in $L_{\rm bol}$ implies a reduction in
the nuclear burning rate as this is the only available
energy source. A reduction by a factor 4 on such short time scales
defies physical principles of thermodynamic cooling and appears impossible
to reproduce with nova evolution models.\\

Meanwhile, in the occultation model for U\,Sco, the observed
blackbody-shaped continuum emission was interpreted by Thomson scattering
in the ambient medium. This, however, would lead to a smaller radius
(that is derived from the normalisation) while $T_{\rm eff}$
would be unchanged. If, however, we are not dealing with Thomson
scattering but significant contributions from Compton scattering, the
photon energies would be
reduced with each scattering event, leading to a softer spectrum that
would yield lower $T_{\rm eff}$. This would lead to an
apparent reduction in flux, and a reduced radius would not be needed.
For photons so well below 511\,keV as in our case, a single Compton
scattering event does reduce the photon energy sufficiently to lead to
such a large reduction in measured effective temperature, thus multiple
scattering events would be needed as found in an optically-thick
photoionized blob. Since treating the radiative transfer with Compton
redistribution is complex we are unable to assess in detail the
viability of Compton scattering but refer to \cite{Compton} for more
work on Compton versus Thomson scattering in White Dwarfs.\\

It is thus not necessary to give up the concept of constant bolometric
luminosity, and it would need to be revisited if the assumption of
Thomson scattering for the spectral analysis of U\,Sco could actually
be replaced by Compton scattering. For U\,Sco, however, we do not
have the original photospheric spectrum and a comparison is not so
straight forward as in the case of V3890\,Sgr.\\

As cautioned before, these conclusions are model-dependent, and the uncertainty
ranges given in Table~\ref{table:rgs_fit_colors} are only statistical
uncertainties, thus with 68\% likelihood, results within these ranges are
found again when repeating the fit with the same data and same model. The
accuracy of the values depends also on quality and completeness of
underlying atomic data and calibration uncertainty in the observations
which are more difficult to quantify. Further, the model assumptions
may be different from reality in which case none of the conclusions
hold. Wrong model assumptions can of course not be quantified.\\

As an example, the intrinsic value of $N_{\rm H}$ was found much lower
in the dip spectrum ($<0.6\times 10^{21}$\,cm$^{-2}$) than in the
bright spectrum ($18\pm 3\times 10^{21}$\,cm$^{-2}$); see Table~\ref{table:rgs_fit_colors}. Lower $N_{\rm H}$
allows more soft emission to pass the interstellar medium which can
then lead to a lower temperature. Fixing $N_{\rm H}$ to the value of
the bright spectrum yields the same or even higher value of $T_{\rm eff}$
for the dip spectrum, then not requiring the conclusion of Compton
rather than Thomson scattering. The result thus depends strongly
on the details of modelling photoelectric absorption.

\section{Summary and Conclusions}

The 3rd recorded outburst of V3890\,Sgr has been observed in X-rays with
\swift, \asat, \nicer, \chandra, and \xmm. All these observatories consistently
found a slowly fading hard collisional emission line spectrum and a highly
variable super-soft source (SSS) atmospheric emission spectrum. All evidence
supports that the collisional emission originates from shocks between the nova
ejecta and the dense wind of the giant stellar companion star. The SSS
spectrum is consistent with emission powered by residual nuclear burning on
the surface of the white dwarf.
Meanwhile, \swift/UVOT and \xmm/OM light curves show a slow, smooth, decline
in UV like in the collisional X-ray emission. The UV emission is thus
unrelated to the SSS emission from the white dwarf, both in terms of direct
emission and inelastic scattering.
The high-resolution \xmm/RGS spectrum revealed a clear reduction of the
shock temperature compared to an earlier \chandra/HETGS observation.\\

The focus of this work lies on the timing and spectral analysis of the
\xmm\ data of the SSS component. The high degree of variability of the SSS
component manifests itself in the \xmm\ X-ray light curves as variations
up to a factor 4 between a bright state (65\% of time) and two low states
(12\% of time), dominated by one deep dip in the middle of the
observation and a likely similar dip that is only seen in egress at the
start of the observation.\\

{\bf A possible 18.1-minute oscillation}: During the bright phases, we
see variations of order 10\% that appear periodic with 18.1-minutes
which are formally statistically significant after the deep dip but with
low duty cycle. This
period was not found in the long-term \asat/SXT data, however, a 10\%
brightness variation is close to the limit of what can be detected
with \asat.
Possible interpretations may be pulsations from the surface of the white
dwarf (possibly related to the rotation of the white dwarf) that are
damped while propagating to the outer layers. If the photospheric radius
has shrunk after recovering from the deep dip, the amplitude of the
damped pulsations, then at lower radius, might have increased to become
detectable.\\
Another interpretation may be the presence of slightly denser (and thus cooler)
co-rotating bulges (e.g., at the magnetic poles) that are buried within
the optically thick envelope until the photospheric radius shrinks to
a critical limit. Again, if the recovery from the deep dip has resulted
in a smaller photospheric radius, one or both co-rotating bulges would have
become visible and thus modulate the light curve with the white dwarf
rotation period.\\

{\bf Eclipses by clumps} are another possibility to explain the high-amplitude
variations. Clump formation is expected in photoionized winds like in
O stars, and a phenomenological eclipse model with seven clumps rotating
on orbits between 5-150 times the radius of the white dwarf and sizes between
0.5-10 times the radius of the white dwarf can reproduce the observed
light curve. In the model, each clump only crosses the line of sight once
as the respective orbital periods are longer than the observation duration.
Since the spectral shape with given energy of the Wien tail dictates that the
central source must be very small, partial eclipses require a certain
fine-tuning, however, the system likely contains a much higher number
than seven clumps, and only a small fraction of them will lead to
partial or total eclipses.\\

{\bf The dip spectrum} is dominated by mildly broadened emission lines
at rest wavelengths, indicating that
all SSS photospheric emission has disappeared, leaving behind 'coronal'
emission originating from the cooler components of the shock plasma and
possibly photoionised thin ejecta. The phenomenon of total disappearance
of SSS emission during short periods of time has been observed before,
e.g. in V4743\,Sgr or RS\,Oph. The steep recovery of emission to pre-dip
levels after such dips supports interpretations of geometrical occultations
compared to such abrupt changes in the nuclear burning rate. In the high-inclination
U\,Sco system, the X-ray spectrum was all the time dominated by such
emission lines and was interpreted as obscuration of the central SSS emission by
the accretion disc or stream. Residual blackbody emission was seen and
interpreted as Thomson scattering, and a weak blackbody component can
also be identified in V3890\,Sgr, although only with a spectral model
(see below).\\

{\bf The photospheric X-ray absorption lines} are resolved in the \xmm/RGS spectrum and
are systematically blue-shifted by $\sim 800-1500$\,km\,s$^{-1}$ and
broadened within the same range. Other novae such as V2491\,Cyg
are  much less homogeneous in these parameters, and this simplifies
the spectral modeling.\\
The measurement of these velocity
values is compromised by the relatively strong 'coronal' background
emission lines seen at rest wavelengths during the dip which then
fill up the red wings of the blue-shifted absorption lines making them
appear narrower than they would leave the photosphere.\\

{\bf The SPEX modelling} results in highly satisfactory reproduction
of the complex SSS RGS spectrum with a relatively low number of components
assuming a blackbody source function that is absorbed by a single
photoionised layer (called {\scriptsize{PION}}
in SPEX); for comparison, for V2491\,Cyg,
three {\scriptsize{PION}} layers had to be assumed. The CNO abundances were found to be
non-solar as expected for nuclearly processed material.
The shock component is also overabundant in N but less than in the
photosphere indicating mixing between CNO-enrichted nova ejecta and
solar-composition material from the companion star.\\

The SPEX model was also used to probe spectral changes with the
variability patterns. In the deep-dip spectrum, a blackbody component
can be detected yielding a similar surface area but lower $T_{\rm eff}$
compared to the bright spectrum. This implies a reduced value of $L_{\rm bol}$
and thus a reduced nuclear burning rate which may be difficult to
achieve on the short time scales of in- and egress of the deep dip.
The SPEX results and the clump occultation model combined lead to the
picture of total occultation of the central source and Compton scattering
of photospheric emission in the ambient, electron-rich medium. 
Compton scattering leads to a softer spectrum, thus lower $T_{\rm eff}$
and reduced brightness without the need of reduced $L_{\rm bol}$.\\
 
The SPEX model also detected a small increase in blue shift of
absorption lines from $870\,\pm\,10$\,km\,s$^{-1}$ before the dip
to $900\,\pm\,10$\,km\,s$^{-1}$ after the dip. This supports interpretations of
a smaller photospheric radius after recuperation from the dip which
would also support both hypotheses for seeing 18.1-minute oscillations
only after the dip. Possible speculation might be that a large occulting
clump has ripped off a small amount of plasma from the
photospheric layers as a result of a grazing passage, or it was ejected
from the optically thick envelope.\\

{\bf Overall Conclusions}\\
The mystery of fast and high-amplitude variations started with
the steep decline in a 2002 \chandra\ observation of V4743\,Sgr,
and culminated with high-amplitude variability in the 2006 \swift\
monitoring observations of RS\,Oph. We have shown that the concept
of constant $L_{\rm bol}$ does not need to be given up as
such steep brightness variations can be caused by occultations while
the reduced photospheric temperature in the scattered light can be
explained by Compton scattering processes.

\begin{acknowledgements}
A.P. Beardmore, J.P. Osborne, K.L. Page
acknowledge support from the UK Space Agency.
M.H. acknowledges support from an ESA fellowship.
AD and PB were supported by the Slovak grant VEGA 1/0408/20, and by the Operational Programme Research and Innovation for the project: Scientific and Research Centre of Excellence SlovakION for Material and Interdisciplinary Research“, code of the project ITMS2014+: 313011W085 co-financed by the European Regional Development Fund.
S. Starrfield gratefully acknowledges partial support
from NSF and NASA grants to ASU.
BV gratefully acknowledges support by the Research Foundation - Flanders (FWO-Vlaanderen, project 11H2121N)
\end{acknowledgements}

\bibliographystyle{aa}
\bibliography{cn,jn,rsoph,astron,v3890sgr}

\end{document}

%% file: tab.tex
N\,{\sc vi}\,$\alpha$ (28.78)&$-887_{-154}^{+156}$&$1350_{-247}^{+296}$&$0.64_{-0.10}^{+0.11}$&0.72&$0.12_{-0.02}^{+0.02}$\\
\hfill {\it add} &$-753_{-313}^{+342}$&$970_{-382}^{+516}$&$1.12_{-0.51}^{+1.06}$&&$0.21_{-0.09}^{+0.21}$\\
\hline
N\,{\sc vi}\,$\beta$ (24.90)&$-2391_{-190}^{+97}$&$591_{-107}^{+174}$&$0.71_{-0.20}^{+0.30}$&0.17&$0.76_{-0.21}^{+0.33}$\\
\hfill {\it add} &$-2573_{-388}^{+1289}$&$1293_{-707}^{+366}$&$1.67_{-1.31}^{+1.20}$&&$1.80_{-1.41}^{+1.29}$\\
\hfill {\it diff} &$-2546_{-391}^{+1264}$&$1306_{-669}^{+373}$&$1.63_{-1.25}^{+1.20}$&&$1.76_{-1.35}^{+1.29}$\\
\hline
N\,{\sc vii}\,$\alpha$ (24.78)&$-2183_{-218}^{+109}$&$1060_{-176}^{+108}$&$1.28_{-0.14}^{+0.08}$&0.83&$0.29_{-0.03}^{+0.02}$\\
\hfill {\it add} &$-2388_{-634}^{+513}$&$805_{-597}^{+414}$&$1.34_{-1.07}^{+1.71}$&&$0.30_{-0.24}^{+0.39}$\\
\hfill {\it diff} &$-2365_{-658}^{+495}$&$816_{-591}^{+412}$&$1.39_{-1.09}^{+1.65}$&&$0.31_{-0.24}^{+0.38}$\\
\hline
N\,{\sc vii}\,$\beta$ (20.90)&$-1168_{-112}^{+128}$&$1437_{-206}^{+286}$&$1.02_{-0.14}^{+0.16}$&0.16&$1.68_{-0.24}^{+0.27}$\\
\hfill {\it add} &$-1177_{-242}^{+267}$&$1781_{-452}^{+766}$&$1.17_{-0.30}^{+0.42}$&&$1.93_{-0.50}^{+0.70}$\\
\hline
N\,{\sc vii}\,$\gamma$ (19.83)&$-938_{-74}^{+7417}$&$1372_{-118}^{+124}$&$0.59_{-0.04}^{+0.05}$&0.06&$2.96_{-0.24}^{+0.24}$\\
\hfill {\it add} &$-909_{-145}^{+144}$&$1264_{-201}^{+235}$&$0.70_{-0.12}^{+0.13}$&&$3.48_{-0.58}^{+0.65}$\\
\hline
N\,{\sc vii}\,$\delta$ (19.36)&$-883_{-81}^{+79}$&$1013_{-121}^{+145}$&$0.44_{-0.05}^{+0.06}$&0.03&$4.83_{-0.55}^{+0.57}$\\
\hfill {\it add} &$-902_{-202}^{+183}$&$1036_{-278}^{+435}$&$0.47_{-0.12}^{+0.12}$&&$5.10_{-1.27}^{+1.38}$\\
\hline
O\,{\sc vii}\,$\alpha$ (21.60)&$-1158_{-70}^{+70}$&$1085.4_{-94}^{+107}$&$1.32_{-0.13}^{+0.15}$&0.71&$0.45_{-0.04}^{+0.05}$\\
\hfill {\it add} &$-831_{-225}^{+229}$&$1322_{-310}^{+476}$&$1.63_{-0.41}^{+0.71}$&&$0.56_{-0.14}^{+0.24}$\\
\hline
O\,{\sc vii}\,$\beta$ (18.63)&$-1381_{-200}^{+192}$&$4342_{-437}^{+495}$&$0.79_{-0.05}^{+0.05}$&0.15&$1.62_{-0.10}^{+0.10}$\\
\hfill {\it add} &$-1595_{-398}^{+465}$&$2964_{-1010}^{+1134}$&$1.38_{-0.40}^{+0.46}$&&$2.84_{-0.84}^{+0.95}$\\
\hfill {\it diff} &$-1415_{-290}^{+295}$&$2597_{-480}^{+610}$&$1.45_{-0.32}^{+0.41}$&&$5.79_{-1.31}^{+1.79}$\\
\hline
O\,{\sc vii}\,$\gamma$ (17.77)&$-1325_{-122}^{+119}$&$865_{-163}^{+201}$&$0.48_{-0.09}^{+0.10}$&0.06&$2.92_{-0.54}^{+0.58}$\\
\hfill {\it add} &$-1344_{-323}^{+354}$&$998_{-395}^{+1034}$&$0.75_{-0.32}^{+0.46}$&&$4.53_{-1.97}^{+2.78}$\\
\hline
O\,{\sc vii}\,$\delta$ (17.39)&$-835_{-119}^{+120}$&$663.23_{-125}^{+149}$&$0.47_{-0.11}^{+0.12}$&0.03&$5.90_{-1.37}^{+1.53}$\\
\hfill {\it add} &$-706_{-252}^{+274}$&$680_{-261}^{+343}$&$1.04_{-0.51}^{+1.00}$&&$13.1_{-6.43}^{+12.6}$\\
\hline
O\,{\sc viii}\,$\alpha$ (18.97)&$-1456_{-119}^{+106}$&$898_{-136}^{+169}$&$0.48_{-0.08}^{+0.08}$&0.83&$0.18_{-0.03}^{+0.04}$\\
\hfill {\it add} &$-1130_{-447}^{+426}$&$1476_{-485}^{+687}$&$1.34_{-0.56}^{+0.82}$&&$0.51_{-0.22}^{+0.32}$\\
\hfill {\it diff} &$-1159_{-294}^{+269}$&$1490_{-341}^{+540}$&$1.35_{-0.39}^{+0.54}$&&$1.03_{-0.31}^{+0.41}$\\
\hline